\documentclass[aps,prd,twocolumn,preprintnumbers,superscriptaddress,dblfloatfix,nofootinbib]{revtex4-1}
\usepackage[utf8]{inputenc}
\usepackage{lmodern}
\usepackage{amsmath,amssymb,mhchem}
\usepackage{graphicx}
\usepackage{url}
\usepackage{color}
\usepackage{enumitem}
\usepackage{subfigure}
\usepackage[dvipsnames]{xcolor}
\usepackage[colorlinks=true,breaklinks=true]{hyperref}
\hypersetup{allcolors=[rgb]{0.0 0.0 0.6},linkcolor=[rgb]{0.75 0.05 0.05}}
\usepackage{bm}
\usepackage{epsfig}
\usepackage{amsmath}
\usepackage{amssymb}
\usepackage{slashed}
\usepackage{color}
\usepackage{accents}
\usepackage{url}
\usepackage[dvipsnames]{xcolor}

\newcommand{\exclude}[1]{}

\begin{document}


\title{Probing dark photons with plasma haloscopes}

\author{Graciela B. Gelmini}\email{gelmini@physics.ucla.edu}
\affiliation{Department of Physics and Astronomy, University of California, Los Angeles \\ Los Angeles, California, 90095-1547, USA}

\author{Alexander J. Millar}\email{alexander.millar@fysik.su.se}
\affiliation{The Oskar Klein Centre for Cosmoparticle Physics, Department of Physics, \\
Stockholm University, AlbaNova, 10691 Stockholm, Sweden} 
\affiliation{Nordita, KTH Royal Institute of Technology and Stockholm University, \\ Roslagstullsbacken 23, 10691 Stockholm, Sweden}

\author{Volodymyr Takhistov}\email{vtakhist@physics.ucla.edu}
\affiliation{Department of Physics and Astronomy, University of California, Los Angeles \\ Los Angeles, California, 90095-1547, USA}

\author{Edoardo Vitagliano}\email{edoardo@physics.ucla.edu}
\affiliation{Department of Physics and Astronomy, University of California, Los Angeles \\ Los Angeles, California, 90095-1547, USA}

\date{\today}

\begin{abstract}
Dark photons (DPs) produced in the early Universe are  well-motivated dark matter (DM) candidates. We show that the recently proposed tunable plasma haloscopes are particularly advantageous for DP searches. While in-medium effects suppress the DP signal in conventional searches, plasma haloscopes make use of metamaterials that enable resonant absorption of the DP by matching its mass to a tunable plasma frequency and thus enable efficient plasmon production. Using thermal field theory, we confirm the in-medium DP absorption rate within the detector. This scheme allows us to competitively explore a significant part of the DP DM parameter space in the DP mass range of $6-400$ $\mu$eV. If a signal is observed, the observation of a daily or annual modulation of the signal would be crucial to clearly identify the signal as due to DP DM and could shed light on the production mechanism.
\end{abstract}
\maketitle
\preprint{NORDITA-2020-054}

\section{introduction}

Uncovering the nature of dark matter (DM), the predominant constituent of matter in the Universe, remains an elusive target and subject of significant scientific efforts~(see e.g. Ref.~\cite{Gelmini:2015zpa} for a review). While a lot of attention has been devoted to exploring DM composed of weakly interacting massive particles (WIMPs), another generic possibility is DM composed of very light ($\ll \mathcal{O}$(eV)) bosonic particles that exhibit classical field-like behavior. 

Dark (hidden) photons (DPs) constitute a prototypical example of spin-1 bosonic light DM (e.g.~\cite{Jaeckel:2010ni,Pospelov:2008jk,An:2013yua}).~Associated with an additional $\mathrm{U}(1)$ dark sector gauge symmetry, they appear in minimal extensions of the Standard Model (SM) and are also well-motivated by fundamental considerations (e.g.~\cite{Goodsell:2009xc}). Analogously to axions~\cite{Preskill:1982cy,Abbott:1982af,Dine:1982ah}, DPs can be readily produced in the early Universe through the misalignment mechanism~\cite{Nelson:2011sf,Arias:2012az, AlonsoAlvarez:2019cgw,Nakayama:2019rhg,Nakayama:2020rka} (other production channels, such as inflationary fluctuations~\cite{Graham:2015rva,Nakai:2020cfw} or decays of pre-cursor fields \cite{Dror:2018pdh,Agrawal:2018vin,Co:2018lka,Bastero-Gil:2018uel, Long:2019lwl}, are also possible). Depending on its production history, the resulting DP vector field may or may not have a preferred direction of alignment at present time within our Galaxy.

A central pillar in experimental efforts to search for the DM are direct detection experiments that attempt to measure the energy deposited within a detector by interactions  of DM in the Milky Way halo passing through it.~Light DM with mass significantly smaller than $\mathcal{O}$(GeV) will not deposit sufficient nuclear recoil energy within conventional ton-scale direct detection experiments (e.g.~\cite{Aprile:2017iyp,Aprile:2018dbl,Akerib:2016vxi}) with keV-level thresholds optimized toward WIMP searches, focusing on electroweak mass scales. Hence, other types of searches are required to explore it. 

For light DM, DPs or axions/axion-like particles (we refer to these generically as ``axions"), sensitivity to  $\mathcal{O}$(eV) mass scales can be gained within large direct detection experiments searching for ionization signals due to DM absorption by bound electrons~\cite{An:2014twa,Bloch:2016sjj}. Detection proposals based on small $\mathcal{O}$(meV) band-gap materials envision good sensitivity for light DM with masses above the gap, allowing for efficient quasi-particle excitation (e.g. phonons), including superconductors~\cite{Hochberg:2015pha,Hochberg:2015fth}, graphene~\cite{Hochberg:2016ntt}, Dirac materials~\cite{Hochberg:2017wce,Coskuner:2019odd}, superfluid helium~\cite{Knapen:2016cue,Schutz:2016tid,Caputo:2019cyg,Acanfora:2019con,Guo:2013dt}, and polar materials~\cite{Knapen:2017ekk}. Axion haloscope experiments~(see~e.g. Ref.~\cite{Irastorza:2018dyq} for a review) can typically also efficiently probe DPs, including setups based on cavity resonators (e.g.~\cite{Sikivie:1983ip,Rybka:2014xca,Woohyun:2016,Goryachev:2017wpw,Alesini:2017ifp,Melcon:2018dba,Melcon:2020xvj}) and dielectrics~\cite{TheMADMAXWorkingGroup:2016hpc,Baryakhtar:2018doz} as well as dish antennas~\cite{Jaeckel:2013sqa,Horns:2012jf,Suzuki:2015sza,Experiment:2017icw,BRASS}. 

In-medium effects within detectors themselves suppress DP interactions and thus present a major limiting factor for experimental searches. If the DM interaction is on resonance with quasi-particle production within the material, such as plasmons, these effects can be mitigated. In a recent study~\cite{Lawson:2019brd}, a novel tunable plasma haloscope for axion detection has been proposed. While the more conventional axion haloscope  cavity searches  for axions coupled to photons rely on resonant axion-photon conversion associated with cavity frequency modes, the plasma haloscope takes advantage of novel metamaterial properties to consider axion-plasmon resonant conversion. The tunable plasma frequency that resonantly matches the axion mass is not directly associated with the dimensions of the detector in plasma haloscopes and allows to efficiently cover a large axion parameter space. 

In this work we show that plasma haloscopes constitute particularly advantageous testing grounds for DPs.  Plasmons have been extensively discussed in relation to dark sectors within astrophysical environments (e.g.~\cite{Raffelt:1996wa,An:2013yfc,An:2013yua,Redondo:2013lna}), the early Universe (e.g.~\cite{Dvorkin:2019zdi}) and more recently in the context of ``inelastic'' nuclear recoils from heavy $\sim$MeV--GeV DM within the more conventional direct detection experiments~\cite{Kurinsky:2020dpb,Kozaczuk:2020uzb}.

While DP-plasmon resonances have been considered in astrophysical environments (e.g.~\cite{An:2013yfc,Redondo:2013lna}), they have not been discussed in laboratory settings. Utilizing a DP-plasmon resonance with a tunable frequency, plasma haloscopes allow to take full advantage of in-medium effects that often plague DP searches, and thus significantly enhance detection sensitivity.

\section{Dark photon in-medium resonance}

The low-energy effective Lagrangian due to the presence of a gauge boson $X$ of a hidden sector U(1) symmetry that kinetically  mixes (e.g.~\cite{Holdom:1985ag,Arias:2012az,An:2014twa,Fabbrichesi:2020wbt}) with the ordinary photon $A$ is
\begin{align}
\mathcal{L} \supset & -\dfrac{1}{4} {F}_{\mu \nu} {F}^{\mu \nu} - \dfrac{1}{4}{X}_{\mu \nu} {X}^{ \mu \nu} +\frac{\sin{\alpha}}{2}F^{\mu\nu}{X}
_{\mu\nu}  \notag \\
& + e J_{\rm EM}^{\mu}{A}_{\mu}+ \dfrac{m_X^2 \cos^2{\alpha}}{2}{X}^{\mu} {X}_{\mu}\, ,
\end{align}
where $F_{\mu \nu}$, $X_{\mu \nu}$ denote the fields strengths of the SM photon and the DP, respectively, $J_{\rm EM}^{\mu}$ is the electromagnetic current, $m_X$ is the DP mass\footnote{The DP mass can be generated via the Higgs or the Stueckelberg mechanisms.} after neglecting terms of order $\alpha^2$  (see below) and $\sin \alpha$ is the kinetic mixing parameter. The kinetic mixing term can be removed by diagonalization through $\tilde{A}={A}\cos{\alpha},  \tilde{X}=X-\sin{\alpha} A$. In the interaction $(\tilde{A},\tilde{X})$ basis, the effective Lagrangian is
\begin{align}\label{lagrangian_interaction}
\mathcal{L} \supset & -\dfrac{1}{4} {\tilde{F}}_{\mu \nu} \tilde{F}^{\mu \nu} - \dfrac{1}{4}\tilde{X}_{\mu \nu} \tilde{X}^{ \mu \nu} + \dfrac{e}{\cos{\alpha}}  J_{\rm EM}^{\mu}\tilde{A}_{\mu} 
  \\
& + \dfrac{m_X^2 \cos^2{\alpha}}{2}\left(\tilde{X}^{\mu} \tilde{X}_{\mu} +2\chi\tilde{X}_\mu \tilde{A}^\mu+ \chi^2\tilde{A}^\mu \tilde{A}_\mu\right)~,  \notag
\end{align}
where $\chi \equiv \tan {\alpha}$ and $\tilde{A}$, $\tilde{X}$ denote the photon produced in electromagnetic interactions and the DP sterile state, respectively. In the interaction basis the electromagnetic coupling gets renormalized to $(e/\cos \alpha)$ and there are $\tilde{A} - \tilde{X}$ photon-DP oscillations due to mass-mixing.

 Neglecting $\mathcal{O}(\chi^2)$ terms and dropping the tilde, we have
\begin{align}\label{lagrangianinteraction}
\mathcal{L} \supset & -\dfrac{1}{4} {{F}}_{\mu \nu} {F}^{\mu \nu} - \dfrac{1}{4}{X}_{\mu \nu} {X}^{ \mu \nu} + e  J_{\rm EM}^{\mu}{A}_{\mu}
\notag \\
& + \dfrac{m_{X}^2}{2}\left({X}^{\mu} {X}_{\mu} +2\chi{X}_\mu {A}^\mu\right) \, ,
\end{align}
which gives the wave equation in vacuum (in momentum space)
\begin{align}
    -K^2 {A}^\nu= \chi m_X^2{X}^\nu \, 
\end{align}
where we defined the four momentum $K=(\omega,\bold{k})$. Here we have used the Fourier expansion for a free field with the energy  $\omega=+\sqrt{|\bold{k}|^2 +m_X^2}$. We treat the fields as complex, $X^\mu_c(t, \bold{x})$, of which the actual fields constitute the real part $X^\mu=$ Re$\{X^\mu_c\}$. As in Ref.~\cite{Knirck:2018knd}, we include a volume $V$ in the definition of the Fourier transform, which will later simplify the definition of the DM density in Eq.~\eqref{density},
\begin{align}
\label{fourier}
    {X_c^\mu}(t,\bold{x})= \sqrt{V} \int \frac{d^3 \bold{k}}{(2\pi)^3} &X^\mu (\bold{k}) e^{-i(\omega t -\bold{k}\bold{x})}\, .
\end{align}
 Our use of a classical field description is justified as the state occupation number is very large~(e.g.~\cite{Jaeckel:2013sqa}).

For a DP that kinetically mixes with the SM photon, effects of in-medium propagation can be significant. In-medium interactions can be accounted for by including a linear response (e.g.~\cite{Raffelt:1996wa,Coskuner:2019odd,Haft:1993jt}), 
\begin{align}
    J^\mu_{\rm ind}=-\Pi^{\mu\nu}{A}_\nu \, ,
\end{align}
where $\Pi^{\mu\nu}$ is a polarization tensor. Hence, the total current coupling to photon is $e J_{\rm EM}^{\mu} + J^\mu_{\rm ind}$. Let us first assume an isotropic and parity invariant medium, and later discuss the extremely anisotropic wire metamaterial scenario.
In the Lorentz gauge, the polarization tensor is described by two polarization functions $\Pi_T$ and $\Pi_L$, for transverse and longitudinal excitations (e.g.~\cite{An:2014twa})
\begin{align}
    \Pi^{\mu\nu}\equiv ie^2\langle J^\mu_{\rm EM}J^\nu_{\rm EM}\rangle&=-\Pi_T (e_+^\mu e_+^{*\nu}+e_-^\mu e_-^{*\nu})-
    \Pi_L e_L^\mu e_L^{*\nu}
    \notag
    \\
    &=\sum_{i=\pm, L} \Pi_i P_i^{\mu\nu}~,
\end{align}
where $e_L$ and $e_{+,-}$ are the longitudinal and the transverse polarization vectors, respectively, and $P_i^{\mu\nu}$ are projectors. Assuming $\bold{k}$ parallel to the $\hat{z}$ axis, the polarization vectors are
\begin{align}
e_L\equiv\frac{(k^2,\omega\bold{k})}{k\sqrt{K^2}} \quad{\rm and}\quad 
e_{+,-}\equiv\frac{1}{\sqrt{2}}(0,\bold{e}_x\pm i \bold{e}_y) \, ~,
\end{align}
where $\bold{e}_x, \bold{e}_y$ are orthogonal  unit vectors perpendicular to the unit vector $\bold{k}/k$.

The wave equation for the interaction eigenstate ${A}=(A^0,\bold{A})$, in the absence of a $J^\mu_{\rm EM}$ reads
\begin{align}
    -K^2 {A}^\nu=-\Pi^{\mu\nu}{A}_\mu+\chi m_X^2{X}^\nu \, .
\end{align}

The relationship between the dielectric tensor $\epsilon$ and the polarization tensor components is given by~\cite{Weldon:1982aq,An:2014twa}
\begin{align}
\epsilon_L=1-\frac{\Pi_L}{\omega^2-k^2} 
\quad {\rm and} \quad
\epsilon_T=1-\frac{\Pi_T}{\omega^2} 
\end{align}
so that for small $k$ (small velocity of the DM) $\epsilon_T=\epsilon_L=\epsilon$.
The dielectric function has real and imaginary parts, just like the polaritazion tensor. The real part is a phase shift in a travelling photon wave function, while the imaginary part is due to absorption of photons travelling in the medium. In the Drude model, typically used to describe metals,  $\epsilon$ can be written as
 \begin{equation}
    \epsilon = 1 - \dfrac{\omega_p^2}{\omega^2 - i \omega \Gamma}\simeq 1-\frac{\omega_p^2}{\omega^2}+i\frac{\Gamma \omega_p^2}{\omega^3}~,
\end{equation}
where $\Gamma$ is the damping within the medium.
For small $k$ we have that  $\omega\simeq m_\chi$ and the zero-components of the fields are negligible,  $A^0\simeq  (\bold{k} \cdot \bold{A} / \omega) \simeq  0$ and $X^0\simeq  0$ (see Appendix of Ref.~\cite{Jaeckel:2013sqa}). Thus, the ordinary  electric field produced by DP (assuming an isotropic medium) is given by
\begin{align}\label{electricfield}
|\bold{E}(\bold{k=0})|=\left|\frac{\chi ~ m_X }{\epsilon}\bold{X}(\bold{k=0})\right| \, .
\end{align}

\section{Output power}
We would like to relate the $\bold{X}$ field in Eq.~\eqref{electricfield} with the DM energy density. First, we note that the detected electric field, as produced by DP in Eq.~\eqref{electricfield}, should be the time derivative of the interaction eigenstate\footnote{We have a different result with respect to the one quoted in Ref.~\cite{Baryakhtar:2018doz} (see the discussion after their Eq.~(9)), in which what is called ``electric field'' is in our case $(1-\epsilon)\bold{E}$.  However, the final results of Ref.~\cite{Baryakhtar:2018doz} do not depend on this expression (nor do ours, as we tune the haloscope to obtain $\epsilon\rightarrow 0$).}.

Another remark is related to the field $\bold{X}$ in Eq.~\eqref{fourier}, which is the sterile interaction eigenstate. In contrast, the DP DM is residing in a vacuum propagation eigenstate (i.e.~a mass eigenstate), after a nontrivial cosmological evolution~\cite{Arias:2012az,McDermott:2019lch,Witte:2020rvb,Caputo:2020bdy,Dubovsky:2015cca}. The vacuum propagation eigenstates $({A}',{X}')$, found by diagonalizing the mass term of Eq.~\eqref{lagrangian_interaction}, can be related to $(A, X)$ via
\begin{align}
\begin{pmatrix}
{A} \\
{X}
\end{pmatrix}=
\begin{pmatrix}
\cos{\alpha} & \sin{\alpha} \\
-\sin{\alpha} & \cos{\alpha}
\end{pmatrix}
\begin{pmatrix}
{A}' \\
{X}'
\end{pmatrix}\simeq
\begin{pmatrix}
1 & \chi \\
-\chi & 1
\end{pmatrix}
\begin{pmatrix}
{A}' \\
{X}'
\end{pmatrix} \, ,
\end{align}
where $\chi$ has been assumed to be small.
We observe that the DM (vacuum propagation eigenstate) consists primarily of the sterile interaction eigenstate, which is the source in Eq.~\eqref{electricfield}. 

At this point, we can relate $\bold{X}$ in Eq.~\eqref{electricfield} with the DM energy density. The Hamiltonian of the DP is that of a massive photon (i.e. the Proca field),
 \begin{align}\label{Hamiltonian}
H_{\rm DP}(t, \bold{x}) =\frac{1}{2} [\bold{E}_X \cdot \bold{E}_X & + \bold{B}_X \cdot \bold{B}_X  
\notag
\\
& + m_X^2 (X^0X^0 + \bold{X} \cdot \bold{X})] ~,
\end{align}
where $\bold{E}_X= \partial_0\bold{X} + \bold{\nabla}X^0$ and $\bold{B}_X= \bold{\nabla} \times \bold{X}$.
Thus, the energy density is~\cite{Jaeckel:2015kea, Arias:2012az}
\begin{align}
\label{density}
\rho &=\frac{1}{V}\int d^3 \bold{x}~ H_{\rm DP}(t, \bold{x})
=\frac{1}{V}\int d^3\bold{x} \ \frac{1}{2}  \partial_0\bold{X} \cdot \partial_0\bold{X}
\notag
\\
& = \int \frac{d^3 \bold{k}}{(2 \pi)^3} \frac{\omega(\bold{k})^2} {2} |\bold{X}(\bold{k})|^2 \, ,
\end{align}
where we used the Fourier expansion in Eq.~\eqref{fourier} and indicated the only term in $H$ whose integral is not zero. 

The space average of the complex DP field is
 \begin{align}
 \label{space-average}
 \langle X_c^\mu(t)\rangle&=\frac{1}{V}\int d^3\bold{x}~ X_c^\mu(t,\bold{x})\notag \\ &=\frac{X^\mu(\bold{k}=0)}{\sqrt{V}}~ e^{-i m_X t}
 \equiv  X_0^\mu~e^{-i m_X t} \, ,
\end{align}
corresponding to a plane wave with frequency $m_X$ whose amplitude
$X_0^\mu=(X_0^0, \bold{X}_0)$ we defined as $X_0^\mu = {X^\mu(\bold{k}=0)}/{\sqrt{V}}$

Equation~\eqref{density}  allow us to make contact between the two different formulations of DP DM as a classical field and as particles, the latter given by the  local velocity distribution of DM particles in the laboratory frame $f_{\rm lab}(\bold{v})$, through 
\begin{align}\label{density-and-f(v)}
\rho= \rho \int d^3 \bold{v} ~ f_{\rm lab}(\bold{v})~. 
\end{align}
Thus, taking $\bold{k}= m_X \bold{v}$ in Eq.~\eqref{density} the DM velocity distribution is identified with~\cite{Knirck:2018knd} 
\begin{equation}\label{f(v)}
f_{\rm lab}(\bold{v})= \frac{m_X^3 \omega^2}{2 (2 \pi)^3 \rho} |\bold{X}(\bold{k})|^2 \, .
\end{equation}

The kinetic energy of DM particles in the dark halo of our Galaxy is of $\mathcal{O}(10^{-6}) m_X$.  Neglecting terms of this order, we can identify $\omega(\bold{k})^2 =m_{\rm X}^2$ in Eq.~\eqref{density}. Then, using the inverse Fourier transform of Eq.~\eqref{fourier} we obtain
\begin{equation}
\label{newrho}
\rho= \dfrac{m_X^2}{2} \langle |\bold{X}_c(t, \bold{x})|^2 \rangle~,
\end{equation}
where the brackets indicate the space average of the square magnitude of the DP field.
Further, neglecting the DM velocity that is of $\mathcal{O}(10^{-3})$, i.e.  using
\begin{align}\label{X(p)}
 X^\mu(\bold{k}) &=\frac{X^\mu(\bold{k}=0)}{V}~ (2 \pi)^3 \delta^3(\bold{k})
= \frac{X^\mu_0}{\sqrt{V}~ }(2 \pi)^3 \delta^3(\bold{k})  
\end{align}
in the Fourier expansion Eq.~\eqref{fourier} (which means
approximating the local DP DM velocity distribution by
$f_{\rm lab}(\bold{v})= (m_X^2/2) \rho |\bold{X}_0|^2 \delta^{(3)}(\bold{v})$),
allows us to identify the DP field with the plane wave in 
Eq.~\eqref{space-average}. Therefore,
 \begin{align}\label{density-X0}
\rho= \frac{m_X^2}{2}~|\langle \bold{X}_c(t)\rangle|^2
=\frac{m_X^2}{2}~ \frac{|\bold{X}(\bold{k}=0)|^2}{V}
= \frac{m_X^2}{2} ~|\bold{X}_0|^2 \, .
\end{align}
This result immediately follows from Eq.~\eqref{newrho} for a field constant in space and can be obtained directly using Eq.~\eqref{X(p)} and $ [(2 \pi)^3 \delta^3(\bold{k})]^2 = V(2 \pi)^3 \delta^3(\bold{k})$ in the integral of Eq.~\eqref{density}.
 
We can also make contact through the expressions we obtained for the DM energy density $\rho$ with the momentum density function $f_X(\bold{k})$ that we employ in Appendix~\ref{alternative}. Writing the energy density as
\begin{align}\label{density-and-f(p)}
\rho = \int \frac{d^3 \bold{k}}{(2 \pi)^3} ~ \omega(\bold{k})~ f_X(\bold{k})~,
\end{align}
 we can use Eq.~\eqref{density} and obtain
 \begin{align}\label{f(p)}
f_X(\bold{k})= \frac{\omega(\bold{k})}{2} |\bold{X}(\bold{k})|^2~.
\end{align}
Alternatively, if we neglect the DM velocity, i.e.~using Eq.~\eqref{X(p)}, we find
 \begin{align}\label{f(p)-with-delta}
f_X(\bold{k})= \frac{(2 \pi)^3}{2} m_X |\bold{X}_0|^2 \delta^{(3)}(\bold{k}) ~.
\end{align}

The power $P$ in a plasma is given by 
\begin{align}
  P = \Gamma U~,
\end{align}
where $U$ is the stored energy. The damping is given by $\Gamma = \omega/Q$, where $Q$ is the quality factor and $(\omega/Q)$ denotes the full-width-half-maximum of the signal. 

For DP, the on resonance output power, when put in a suggestive form reminiscent of cavity haloscopes~\cite{Arias:2012az}, is thus given by  
\begin{align} \label{eq:power2}
    P_{\rm out} &=\kappa~\Gamma~ \dfrac{1}{4}\int\Big(\dfrac{\partial(\epsilon \omega)}{\partial\omega}|\mathbf{E}|^2+|\mathbf{B}|^2\Big)dV \notag\\
    & = \kappa ~ \chi^2 \rho~ m_X Q V_d ~ \mathcal{G}\, ,
\end{align}
where in the first line we used the definition of energy density of the electromagnetic field within a dispersive medium~\cite{landau2013electrodynamics} and in the second line we used Eq.~\eqref{density-X0} and defined the ``geometric factor'' 
\begin{equation} \label{eq:geomf}
    \mathcal{G} =  \dfrac{|\epsilon|^2}{m_X^2 \chi^2|\bold{X}_0|^2 V_d} \dfrac{1}{2}\int\Big(\dfrac{\partial(\epsilon \omega)}{\partial\omega}|\mathbf{E}|^2+|\mathbf{B}|^2\Big)dV \, ,
\end{equation}
which is typically of $\mathcal{O}(1)$. Note that the $E$-field structure is the same as in the case of the axion (Ref.~\cite{Lawson:2019brd}). Here, $\rho = 0.45$~GeV/cm$^3$ is the local DM density (the value assumed in e.g. Ref.~\cite{Asztalos:2009yp} and compatible with observations~\cite{deSalas:2019rdi, Benito:2019ngh,Read:2014qva}), $V_d$ is the fiducial volume of the detector, and $\kappa$ is the signal coupling efficiency factor.

In Appendix~\ref{alternative}, we employ the machinery of thermal field theory to directly compute the power absorbed by the detector in the ``propagator approach''~\cite{Redondo:2013lna}. Our results confirm the expressions obtained with classical electrodynamics in the main text, as well as the expressions obtained with quantum field theory (with in-medium corrections), as used in e.g. Refs.~\cite{An:2014twa,Hochberg:2017wce}.

\section{Thin wire metamaterials}
Recent advances in material science have devoted significant attention to metamaterials, composite materials with periodically or randomly distributed artificial micro-structure with size and spacing smaller than the wavelength of interest.~Their unique properties can be exploited for new physics searches.

We briefly describe wire metamaterials, which have recently been proposed for the realization of tunable plasma haloscopes~\cite{Lawson:2019brd}. As demonstrated in Ref.~\cite{Pendry:1998}, an extended network of thin wires comprises a composite dielectric that behaves like a metal with a plasma frequency in the GHz range. We envision a configuration based on a system of aligned wires in the $\hat{z}$-direction. Since such a system constitutes an extremely anisotropic medium, the longitudinal and transverse modes of polarization tensor are mixed. This can be encoded through a matrix~(see e.g. Ref.~\cite{Coskuner:2019odd})
\begin{align}
    \mathcal{K}_{AB}\equiv e_\mu^A \Pi^{\mu\nu}e_\nu^B~,
\end{align}
where the $A$ and $B$ indices run over the longitudinal and transverse polarization vectors. A non-magnetic material has a symmetric polarization tensor and we can thus diagonalize $\mathcal{K}$. In the $ {z}$-direction of the wires the effective dielectric constant can be described by the standard Drude model for metals~\cite{Pendry:1998,Belov:2003}, with eigenvalues
\begin{align}
    \epsilon \equiv \epsilon_z = 1 - \frac{\omega_p^2}{\omega^2 - i \omega \Gamma} \quad \rm{and} \quad \epsilon_{\perp z}=1 \, .
\end{align}
 For a rectangular array of wires, with a wire radius $d$ and an inter-wire spacing in each direction perpendicular to the wires of $a$ and $b$, the plasma frequency  is given by~\cite{Belov:2003} 
\begin{equation}
\label{eq:pasmafreq}
    \omega_p^2 = \dfrac{2\pi/s^2}{\log\left(\dfrac{s}{2\pi d} \right)+F(r)}~, 
\end{equation}
where $s=\sqrt{ab}$, $r=a/b$ and
\begin{equation}
    F(r)=-\frac12 \log r+\sum_{n=1}^\infty\left(\frac{\coth(\pi n r)-1}{n} \right)+\frac{\pi r}{6} \, ,
\end{equation}
allowing for $\omega_p \sim$ GHz when $s \sim$ cm spacing.
Since the plasma frequency is predominantly determined by the inter-wire spacing, this setup allows to realize a tunable plasma haloscope.
The resulting electric field is thus
\begin{align}
|\bold{E}|=\left|\frac{\chi m_X}{\epsilon} \bold{X} \cos{\theta}\right| \, ,
\end{align}
where $\theta$ is
\begin{align}
\label{cos} 
 \cos{\theta}=\frac{|\bold{\hat{z}}\cdot \bold{X}|}{|\bold{X}|}\, .
\end{align}
Envisioning a finite cylindrical experimental configuration, the discussion of bounded plasma solutions for the field propagation in plasma haloscopes can be found in Ref.~\cite{Lawson:2019brd}.

\begin{figure*}[tb]
\centering
\begin{subfigure} 
 \centering
 \includegraphics[width=0.45\linewidth,trim={0 0 0  0},clip]{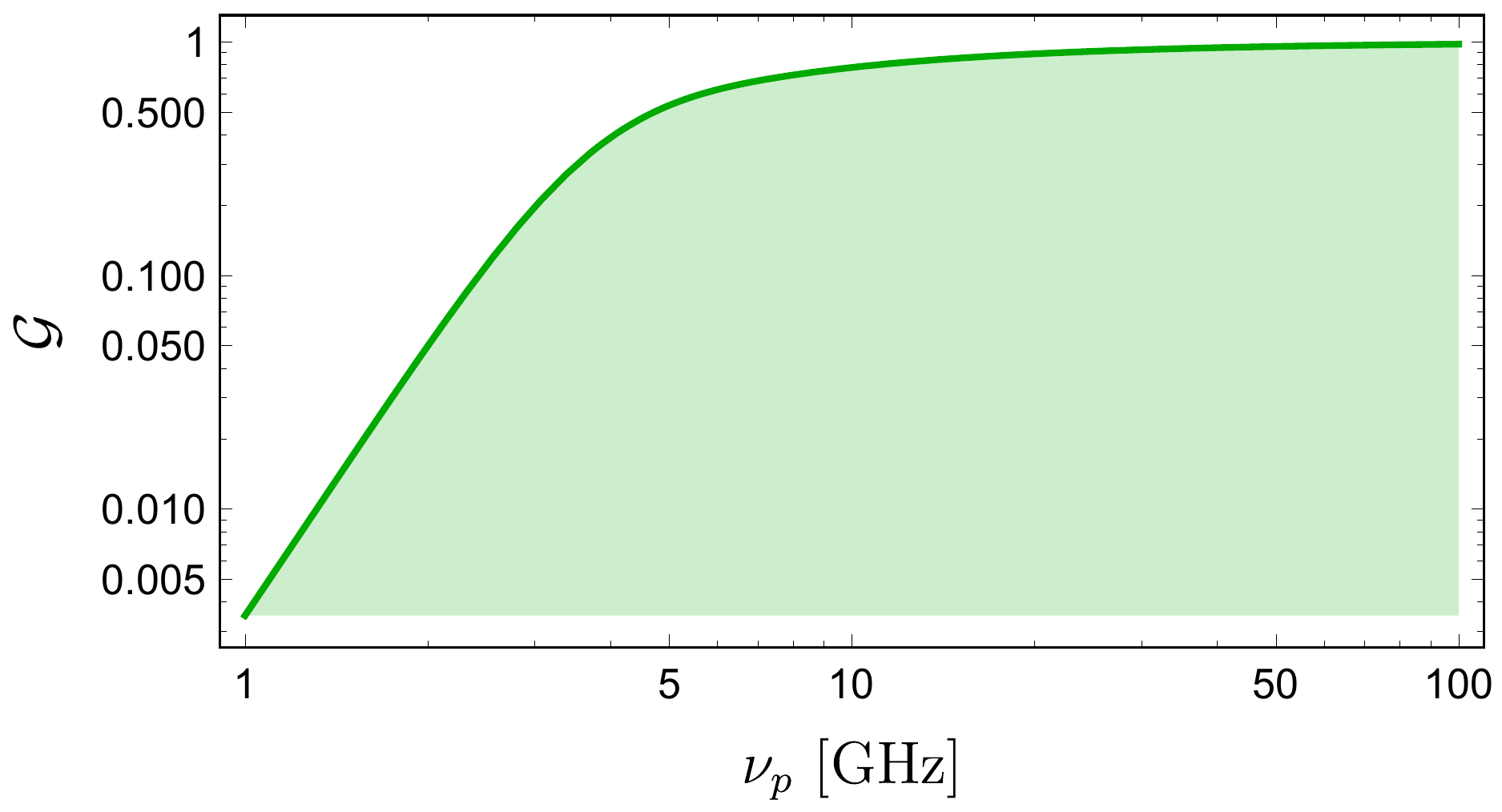} 
\end{subfigure}
\begin{subfigure} 
 \centering
 \includegraphics[width=0.45\linewidth,trim={0 0 0 0},clip]{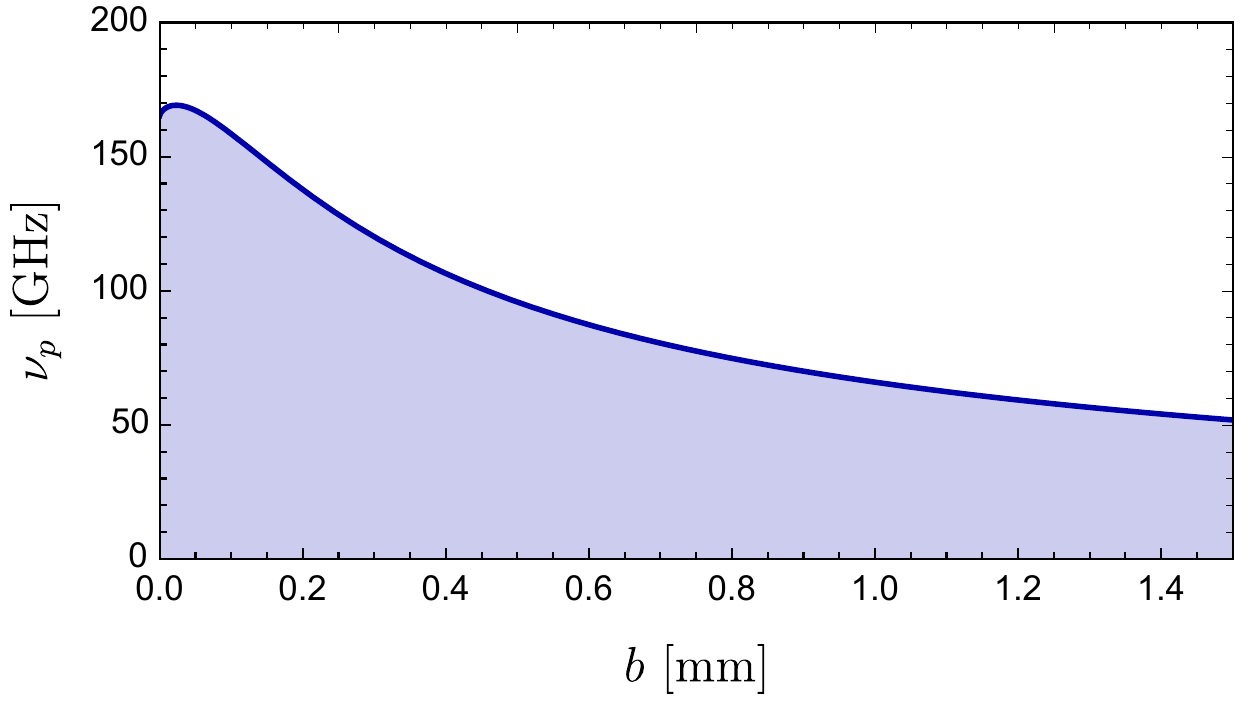} 
\end{subfigure}
\caption{[Left] Geometric factor $\mathcal{G}$ of Eq.~\eqref{eq:geomf}, obtained using finite plasma field description within a wave-guide as derived in Ref.~\cite{Lawson:2019brd}, describing a system of radius 30~cm with a quality factor $Q=100$ as a function of plasma frequency $\nu_p$. [Right] Plasma frequency  $\nu_p=\omega_p/(2\pi)$, given in Eq.~\eqref{eq:pasmafreq}, of a wire metamaterial configuration composed of aligned $10~\mu$m copper wires as a function of inter-wire spacing in orthogonal directions perpendicular to the wires, $a = 1$~mm and $b$. As the spacing $b\to 0$, the plasma frequency saturates to a maximal value. This indicates both the maximum reachable plasma frequency as well as tunability of a given system.}
\label{fig:tuning}
\end{figure*}

Unlike axions, DPs can have electric fields in directions other than the $\hat{z}$-direction of wire alignment. If the boundary of the plasma is a conducting cylinder, for electric fields polarized in the non-axial directions the device behaves like a resonant cavity. Thus, in principle, DPs can excite transverse electric modes (i.e.~modes with $E_z=0$) if the mode frequency matches the DP mass. However, as the DP interacts with cavities analogously to the axion, the rate is proportional to the overlap integral of the DP wave function and the cavity mode. In the case of transverse electric modes in a cylinder, this overlap is zero~\cite{Stern:2015kzo}. Thus, we only consider $E_z$.

\section{Experimental setup}

We now discuss a setup configuration for DP detection with a plasma haloscope. The output power is immediately read from Eq.~\eqref{eq:power2} as
\begin{equation} \label{eq:power}
    P_{\rm out} = \kappa \chi^2   \rho m_{\rm X} Q
V_d ~\mathcal{G} \cos^2\theta~ \, .
\end{equation}
For an unknown DP field direction at the experiment, our limits are obtained by averaging over all the possible directions, with $\langle \cos^2\theta\rangle = 1/3$. Such an average occurs when the polarization of the DP fluctuates over short timescales, so that a measurement samples many different polarizations. In the case of a fixed polarization, for a measurement longer than a day the signal will still be averaged over to give an ${\cal O}(1)$ number, but the exact average will depend on the angle of the DP with the Earth and the experiment's location on Earth.

For simplicity we will follow the specifications of Ref.~\cite{Lawson:2019brd}. We consider a cylindrical structure enclosing copper wires with diameter of $\sim10~\mu$m, which are readily commercially available. We take $Q = 10^2$, $\kappa = 0.5$\footnote{This corresponds to a critically coupled system where the power extracted as signal is equal to other losses.}.

Unlike the axion case~\cite{Lawson:2019brd}, a magnetic field is not required for DP search and hence we are not restricted by the geometric considerations related to the magnet bore size. However, such a device will still need to be cooled, which limits the volume. For a quantum limited detection at low frequencies a dilution refrigerator is required~\cite{Zhong:2018rsr}, which is unlikely to be possible for $V > {\cal O}({\rm m}^3)$. ADMX, which has a cavity volume of 136~L, has the largest dilution refrigerator currently used in axion detection~\cite{Braine:2019fqb}. However, one could also envision a larger, warmer system aiming at lower frequencies. Another limitation for considered volume is that heavier DPs with large de Broglie wavelength $\lambda_{\rm DB}$ will lose coherence over the experiment's dimensions, which both reduces the expected signal and increases the signal's sensitivity to the (unknown) DP velocity dispersion, similar to cavities and dielectric haloscopes~\cite{Knirck:2018knd}.  Hence, each dimension of the experiment should not exceed $\mathcal{O}(10)\%$ of the DP's de Broglie wavelength.
 
Employing the physical dimensions for the experiment as proposed in Ref.~\cite{Lawson:2019brd}, we consider a cylinder of volume $V=0.8~{\rm m}^3$ and diameter 60~cm. For a lower limit on detection frequency, we envision a minimum of 300 wires, which in a 60~cm diameter cylinder gives a wire spacing of 3~cm\footnote{In Ref.~\cite{Lawson:2019brd} there was an error in converting plasma frequencies to physical wire spacings, leading to the frequency associated with a given spacing being overestimated by a factor of $2\pi$. However, as the discussed frequency limits are based on general engineering considerations and not a specific realization of the experiment, this is not very significant.}. The lower limit on the number of wires stems from the requirement that the medium behaves as a plasma, which was confirmed to occur within a square array of $\gtrsim 200-400$ wires~\cite{doi:10.1063/1.1513663}. From Eq.~\eqref{eq:pasmafreq}, this translates into a lower limit on detection sensitivity to plasma frequency $\nu_p = 1.5$~GHz, corresponding to a DP mass of $m_{\rm DP} \simeq 6$~$\mu$eV. These considerations also applied to Ref.~\cite{Lawson:2019brd}, signifying that plasma haloscopes are more sensitive to light DM at smaller mass range than originally thought. Note that when the wavelength of light in the medium becomes larger than the experiment the power becomes suppressed~\cite{Ouellet:2018nfr}. For us, this translates into geometry factor approaching zero as $R/\sqrt{\epsilon_z}\omega\to 0$, where $R$ is the radius of the cylinder. We illustrate this suppression explicitly in the left panel of Fig.~\ref{fig:tuning}, where at $1.5~$GHz ${\cal G}= 1.7\times 10^{-2}$ - leading to a hundredfold suppression of the power. Thus, if even lower frequencies are desired, one must increase the radius of the experiment, at the possible expense of a sub Kelvin cooling system.  An alternative could be to operate slightly off the plasma frequency, thus modifying the wavelength to match a resonant mode of the cavity.

For considerations of a (soft) upper limit on detection frequency, we note that mechanically tuning such a device would prove very challenging for sub-mm wire spacings. However, as can be seen from Eq.~\eqref{eq:pasmafreq}, only one direction needs to be tuned in order to change $\omega_p$~\cite{Lawson:2019brd}. Thus, one can compensate for limitations in the tuning direction by manufacturing wires to be more densely packed in an orthogonal direction. Hence, if the closest possible spacing is $a\sim 1~{\rm mm}$, one could still achieve a plasma frequency of $\nu_p = 100$ GHz with an orthogonal-direction wire spacing of $b\sim 0.5~{\rm mm}$. We note that employing an arbitrarily small spacing in the direction that is not being tuned would not prove beneficial. To illustrate this, we depict in the right panel of Fig.~\ref{fig:tuning} the plasma frequency of Eq.~\eqref{eq:pasmafreq} as a function of spacing $b$ for a set of $10~\mu {\rm m}$ wires of spacing $a=1~{\rm mm}$. We observe that $\omega_p$ is saturated as $b\to 0$ (though the formalism breaks down when the wire radius becomes comparable to the spacing). Hence, if one is restricted to mechanical tuning, the minimum spacing allowed in the tuning direction determines the maximal plasma frequency. To get to $100~$GHz would require a minimum spacing $\lesssim 1.5~$mm. However, we note that wire metamaterials can be tuned via the Josephson effect~\cite{doi:10.1063/1.5126963}, which would evade such constraints.   

\section{signal detection}

We envision detecting the signal via an antenna coupled to the system. While the full design of such an antenna would be the subject of a detailed technical proposal, if multiple elements are needed to extract power from the full volume a summing network could be employed, as was proposed for axion detection in Ref.~\cite{Kuo:2019cps}. As the refractive index is almost purely imaginary on resonance, waves decay over a distance $1/\sqrt{\epsilon_z}\omega$ in transverse (non-z) directions. Thus, systems with $R\gg 1/\sqrt{\epsilon_z}\omega$ require some antenna elements to be placed inside the medium for a full readout. Under these assumptions, we can obtain the scan rate via Dicke's radiometer formula for signal-to-noise ratio $(S/N)$~(see discussion in e.g. Ref.~\cite{Chaudhuri:2018rqn})
\begin{equation} \label{eq:signalnoise}
\dfrac{S}{N} = \dfrac{P}{T_{\rm sys}}\sqrt{\dfrac{\Delta t}{\Delta \nu_{\rm DP}}}~,
\end{equation}
where $T_{\rm sys}$ is the system noise temperature, $\Delta \nu_{\rm DP} \simeq 10^{-6} \nu_{\rm DP}$ is the DP signal line width, $\Delta t$ is the measurement time that covers a frequency range $\sim Q/\omega$. Here, frequency is $\nu_{\rm DP} = \omega/ (2 \pi)$, where $\omega = m_{\rm DP}$ on resonance. We consider quantum limited detection\footnote{At low frequencies ($\lesssim 10~$GHz) detection near quantum limit has been demonstrated with Josephson parametric amplifiers operating at low temperatures~\cite{Zhong:2018rsr}.}, taking $T_{\rm sys} = m_{\rm DP}$.  

For our parameters, the power at a given frequency is 
\begin{align}
    P_{\rm out}=\, &1.1\times 10^{-22} {\rm W}~\Big(\dfrac{\kappa}{0.5}\Big)\Big(\dfrac{{\cal G}}{1}\Big)\left (\frac{\chi}{10^{-15}}\right)^2\Big(\frac{Q}{100}\Big) \notag\\
    &\times\Big(\frac{V_d}{0.8\,{\rm m}^3}\Big)\Big(\frac{\nu}{10\,{\rm GHz}}\Big)\Big(\frac{\rho}{0.45\,{\rm GeV/cm^3}}\Big)\,.
\end{align}
At higher masses, this equation is modified by the restriction that experimental size does not exceed $\sim 30\%$ of $\lambda_{\rm DB}$. As we are considering a narrow aspect ratio cylindrical cavity whose height $h$ exceeds the diameter, we impose $h < 0.3~\lambda_{\rm DB}$. At lower frequencies, the geometric factor~$\mathcal{G}$ of Eq.~\eqref{eq:geomf}, displayed on Fig.~\ref{fig:tuning}, is suppressed due to the finite size of the wave-guide~\cite{Lawson:2019brd}.

We can obtain a rough estimate of the experimental livetime required to scan over a region of parameter space by treating the resonance as a rectangular spectrum with a width given by the full width half maximum ($\omega/Q$) and height given by the half maximum ($P_{\rm out}/2$). This estimate assumes that a relatively rapid frequency tuning ($\ll \Delta t$) without disturbing the system is possible, although these effects could be included~\cite{Millar:2016cjp}. Given that a low $Q$ is assumed, each measurement is relatively long, so this is not a significant restriction on the tuning time. Thus, integrating Eq.~\eqref{eq:signalnoise}, the total scanning time between frequencies $\nu_1$ and $\nu_2$ is given by
\begin{equation}
    t_{\rm scan} =4\times10^{-6}Q\int_{\nu_1}^{\nu_2}d\nu \left(\frac{S}{N}\right)^2\left(\frac{T_{\rm sys}}{P_{\rm out}}\right)^2\,.
\end{equation}
Requiring a signal-to-noise ratio of $(S/N) \geq 3$, a kinetic mixing parameter value down to $\chi \simeq 7\times10^{-16}$ can be probed across our whole parameter range of $6-400~\mu$eV within $\sim 5$ years of experimental livetime.

\begin{figure}[tb]
  \includegraphics[width=0.95\linewidth]{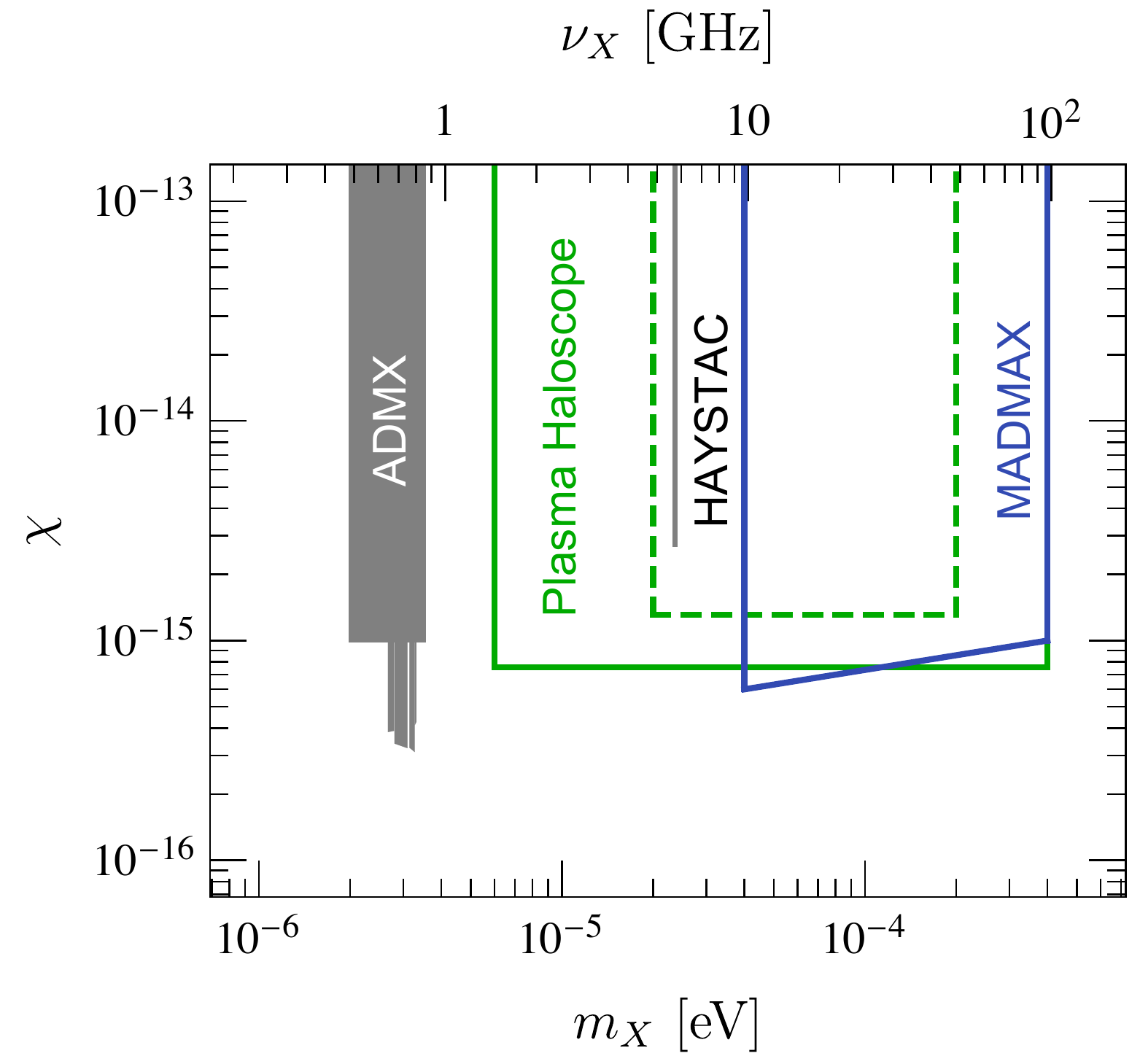}
\caption{Projected sensitivity reach of plasma haloscopes for DP searches assuming an $0.8~{\rm m}^3$ volume of plasma with a quality factor $Q=100$. We assume a 5 year livetime, and display both the optimistic quantum limited (solid green) as well as $8~$K noise temperature (dashed green) cases. Existing limits (dark gray) from haloscope cavity searches of ADMX~\cite{Asztalos:2001jk,PhysRevD.69.011101,Asztalos:2009yp,Du:2018uak} and HAYSTAC~\cite{Brubaker:2016ktl} are shown. We display the projected DP reach of MADMAX (blue) dielectric haloscope, which assumes a constant $8~$K noise temperature~\cite{Brun:2019lyf}.
}
\label{fig:DPlimits}
\end{figure}

For the above input parameters and experimental livetime of 5 years, in Fig.~\ref{fig:DPlimits} we display the projected sensitivity for DP parameter range in green along with existing constraints\footnote{Dish antenna experiments~\cite{Horns:2012jf}, such as SHUKET~\cite{Brun:2019kak} and Tokyo (University of Tokyo)~\cite{Suzuki:2015vka},
also probe this mass range. However, they are more suited for broadband searches at larger values of $\chi$ and their limits are less sensitive than the parameter range displayed.}. We note that while axion haloscope limits are often converted in the literature into limits on DPs in the vein of Ref.~\cite{Arias:2012az}, many of the published limits have employed magnetic field as a veto on signals (e.g.~Refs.~\cite{Wuensch:1989sa,DePanfilis:1987dk}). Thus, any potential DP signals were rejected, signifying that one cannot directly reinterpret such studies as DP searches. In Ref.~\cite{Braine:2019fqb} such a procedure was also used, however only injected simulation candidate events were present before this was done. Hence, while it is uncertain if a DP would have been observed if there had been a candidate event, one can use the lack of candidates as a limit. In the cases which did not explicitly employ such a procedure, we assume that no veto was employed. If the analysis is done allowing for DP signals, cavity searches such as ADMX~\cite{Rybka:2014xca}, HAYSTAC~\cite{Brubaker:2016ktl}, CULTASK~\cite{Woohyun:2016}, ORGAN~\cite{Goryachev:2017wpw}, KLASH~\cite{Alesini:2017ifp} and RADES~\cite{Melcon:2018dba,Melcon:2020xvj} are potentially very sensitive to DPs, particularly for $\nu\lesssim 10~$Ghz. As multiple axion haloscope searches perform such vetoing, we only display published DP projected sensitivities along with existing limits, the most relevant being dielectric haloscopes like MADMAX~\cite{TheMADMAXWorkingGroup:2016hpc,Brun:2019lyf}. With a small enough mass-gap, Dirac materials could also be sensitive in the range of our parameter space of interest~\cite{Hochberg:2017wce}.

To plot limits, we have assumed that the DP field direction changes relatively rapidly, so the averaged angle can be used for all experiments. As individual measurements are long, plasma and dielectric haloscopes would obtain similar results for a fixed DP direction. However, cavity haloscopes generally measure for very short periods, and so would only sample a single value of $\cos^2\theta$ in each measurement. Further, usually multiple measurements, presumably occurring at different times of the day, are combined to get the final limits. To get a limit on a time varying signal would require time and location information for each measurement when combined, in order to take into account signals potentially vanishing at different times of the day. As detailed timing information is usually not reported, it is non-trivial to rigorously turn existing limits into DP limits in such a scenario. 

While for axions both quantum limited and constant noise temperatures were explored in Ref.~\cite{TheMADMAXWorkingGroup:2016hpc}, the limits plotted for DP in Ref.~\cite{Brun:2019lyf} assumed a more modest noise temperature of 8~K. Thus for a direct comparison we plot also a more conservative system with $T_{\rm sys}=8~$K (dashed green line). We note, however, that as near quantum limited detection is currently available below $\sim 10$~GHz, such a plot is somewhat pessimistic in the low mass regime\footnote{Recently an alternatively designed dielectric haloscope, DALI~\cite{DeMiguel-Hernandez:2020mon}, was proposed. However, Ref.~\cite{DeMiguel-Hernandez:2020mon} neglected the quantum limit of linear amplification. HAYSTAC is exploring using squeezed states with Josephson parametric amplifiers at lower frequencies to evade the quantum limit~\cite{Droster:2019fur}, however Ref.~\cite{DeMiguel-Hernandez:2020mon} assumed linear amplifiers would be used, desiring commercially available technology.}.  
As shown in Fig.~\ref{fig:DPlimits} even with conservative parameters plasma haloscopes are capable of searching a large fraction of the DP parameter space, being complementary to dielectric haloscopes like MADMAX. In particular, if both types of experiments use mechanical tuning, the smaller spacings between elements in plasma haloscopes for a given frequency make it more suitable for somewhat lower frequencies.

\section{signal Modulation}

\begin{figure}
\hspace{-10pt}
  \includegraphics[width=1\linewidth]{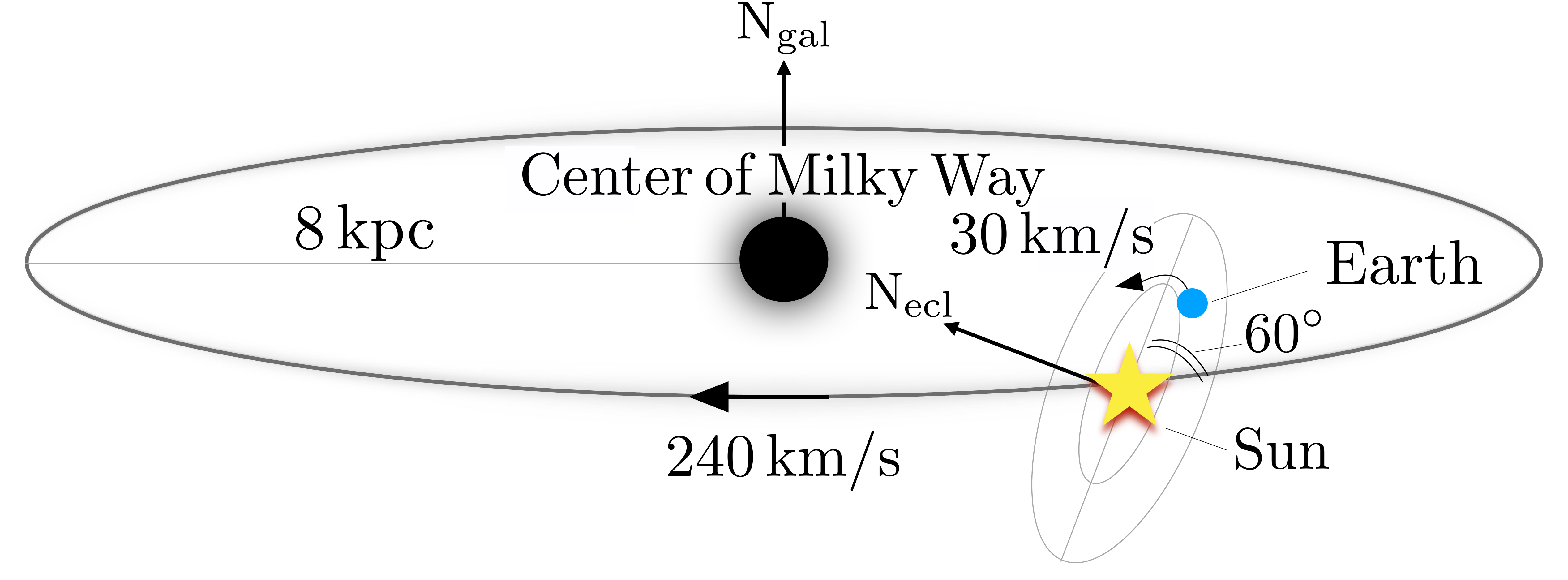}
\caption{\label{sunearthfig} Out of scale diagram of Earth's motion with respect to the Galaxy, due to both the orbital and the Solar System motions. We also show the direction of the north galactic pole $\mathrm{N}_\mathrm{gal}$ and the north ecliptic pole $\mathrm{N}_\mathrm{ecl}$.
}
\end{figure}

Due to Earth's motion with respect to the Galaxy, shown schematically in Fig.~\ref{sunearthfig}, there is a time modulation of a DM signal, since 
the DM distribution is constant in time in the Galactic rest frame during the duration of any experiment (although the local DM distribution at the experiment may change if it enters and/or exists a small enough DM lump or stream, as we comment below).  
Equation~\eqref{f(v)},  where $\bold{v}$ is the velocity of a DM particle in the detector's rest frame, relates the classical field and the particle formulations of DP DM. Aside from the velocity and location of the detector, the orientation of the direction of the wires $\bold{\hat{z}}$ in Eq.~\eqref{cos} also changes periodically in the Galactic rest frame.
 
This leads to three distinct time modulation effects (see e.g.~\cite{Chaudhuri:2018rqn}): (1) the periodic change of the detector speed with respect to the Galaxy, (2) a possible change in DM density during data taking, and (3), the daily change in orientation of the detector with respect to the Galaxy (due to Earth's rotation about itself). This last effect occurs only for DP, but not axions, in some DM generation models and makes use of the strong directionality of the detector we consider.

The time modulation effects due to the periodic change in the velocity of the detector in the Galactic rest frame are very different when the DM particle is absorbed compared to when it scatters off and deposits only a portion of its kinetic energy within the detector. 

When the DM particle is absorbed, the energy deposited in the detector is the particle mass $m_X$ plus its kinetic energy of $\mathcal{O}(10^{-6})m_X$, since the characteristic DM speed is $\mathcal{O}(10^{-3})c$. The DM signal obtains a finite width due to the spread of the kinetic energy.  The peak frequency of the DM line is at $\omega_{\rm peak}= m_X (1+ v_{\rm Sun}^2/2) $, where $v_{\rm Sun}$, the speed of the Sun with respect to the Galaxy,  is about 240 km/s (see e.g.~\cite{Benito:2019ngh} for a discussion of the uncertainties in this speed). 
The orbital motion of Earth,  with speed close to 30 km/s,  
produces an annual modulation of amplitude $\Delta v$ of the detector's speed with respect to the Galaxy, 
 and thus a periodic shift $(v_{\rm Sun}\Delta v)\omega_{\rm peak}$ of $\mathcal{O} (10^{-7})\omega_{\rm peak}$ in the DM line peak frequency. Hence, detecting this modulation would require an experimental energy resolution of $\mathcal{O}(10^{-7})m_X$. 

The existence of a escape speed $v_{\rm esc}$ of about 550 km/s from the Galaxy at Earth's location introduces a high frequency cutoff $\omega_{\rm max}= m_X (1+ (v_{\rm esc} + v_{\rm Sun})^2/2) $ of the DM line. The escape speed has been measured by the RAVE and GAIA surveys, with values between 480 km/s and 640~km/s~\cite{Piffl_2014,
Monari_2018, Deason_2019} with respect to the Galaxy (considering the 90\% C.L. intervals of each measurement). The cutoff could be difficult to measure due to the fast decrease of the DM speed distribution with increasing speed, but it is also annually modulated with an amplitude $[(v_{\rm esc}+ v_{\rm Sun})\Delta v]\omega_{\rm max}$ of $\mathcal{O} (10^{-6})\omega_{\rm max}$.

The daily rotation of Earth around itself, whose surface speed at the equator is $\mathcal{O} (10^{-6})c$,  would induce an even  smaller daily modulation of the peak frequency $\omega_{\rm peak}$ (and also of $\omega_{\rm max}$). This would consequently require an even finer energy resolution of $\mathcal{O}(10^{-9}) m_X$. 

The extremely good energy resolution required to detect the annual modulation of DM line peak, and possibly its daily modulation,  could be achievable with a long enough measurement time once a signal has been observed. The potential detection of the time modulation of the peak frequency, as well as other features of the DM line shape, due e.g. to Sun's gravitational lensing, DM streams, or a dark disk, have been studied at length, primarily in the context of axion detection, see e.g.~\cite{Ling_2004, Vergados_2017, OHare:2017yze,    Foster:2017hbq, OHare:2018trr, Knirck:2018knd, Chaudhuri:2018rqn}.
 
A seemingly erratic time modulation in the amplitude of the signal could occur due to the existence of pervasive very dense and small (relative to the size of the Solar System) clumps of the DM. These clumps are predicted if DP DM is produced due to  fluctuations of the vector field during inflation~\cite{Graham:2015rva}, as described in Ref.~\cite{Chaudhuri:2018rqn}, e.g. if the experiment passes through a clump about every day and spends several seconds within it.

An important daily modulation of the DP signal  could arise due to the strong dependence of the output power of Eq.~\eqref{eq:power} on the orientation of the DP vector field {\bf{X}} with respect to the direction of the wires in the experiment,
\begin{equation}
    P_{\rm out}= P_{\rm out}^{\rm MAX}\cos^2{\theta}\, .
\end{equation}
A daily modulation would only arise if the direction of the DP field is fixed in a large enough region of the Galaxy near Earth, such that the experiment spends  at least several days within it. This could be possible if the DP DM is produced through a misalignment mechanism after inflation, depending on details of structure formation in the Universe (see e.g. Ref.~\cite{Arias:2012az}). 
The possibility of observing a directional modulation has been studied for other detectors for DP detection (see e.g. Ref.~\cite{Jaeckel:2015kea}) and also for axion detection (see e.g. Ref.~\cite{Knirck:2018knd}). 
In the most favorable case in which our detector is located at 45$^o$ latitude and the direction of {\bf{X}} coincides with the direction of the wires, assumed to be in vertical position in the detector,  at one time of the day (so that $\cos{\theta}=1$ at this instant) the daily modulation would be maximal.  The output power would be maximal $P_{\rm out}=P_{\rm out}^{\rm MAX}$ at one time of the day, and 12 hours later (when the azimuth at the location of the experiment would be at 90$^o$ with respect to the original position, thus $\cos{\theta}=0$) the power would be $P_{\rm out}=0$. Thus, in this most favorable case, the daily modulation of the output power would be 100\%. Even just splitting the daily observation into two equal time interval bins in this most favorable case, the total signal in the half day centered at the instant of maximum power would contain  92.4\% of the daily signal (and the other half just the remaining 7.6\%), which would make for a very distinguishable daily modulation. 
Instead, in the worst case scenario, if the {\bf{X}} direction coincides with Earth's rotation axis, $\cos{\theta}$ would be a constant (and related to the latitude of the experiment, always assuming vertical wires) and the daily modulation would be zero. 

The observation of this directional daily modulation of the signal would not only allow to reconstruct the direction of the field {\bf{X}} with respect to the detector, and thus with respect to the Galaxy (any direction in the experiment frame can be easily expressed in the Galactic rest frame  and vice versa --- see e.g. Appendix~A of Ref.~\cite{Bozorgnia:2011tk}, and Ref.~\cite{Mayet:2016zxu}), but could also point toward the DP DM production through the misalignment mechanism.

If the DP DM is instead produced through quantum fluctuations of the vector field during inflation, the direction of the field could change many times during the duration of the experiment, and even during a day. As already mentioned in Ref.~\cite{Chaudhuri:2018rqn}, in this latter  case an optimal setup using simultaneously multiple experiments aligned in different directions would be require to determine the instantaneous direction of $\bold{X}$.

If a signal is observed, the observation of a daily or annual modulation of the signal would be crucial to clearly identify the signal as due to DP DM.

\section{conclusions}

In this work we have shown that recently proposed plasma haloscopes are particularly well suited for DP DM searches.  Plasma haloscopes take advantage of in-medium effects, which suppress the signal in conventional searches. Using description based on thermal field theory as well as classical electrodynamics, we have confirmed the DP absorption rate at the experiment.
By employing metamaterials, the plasma frequency in plasma haloscopes can be tuned to match the DP mass, which allows to competitively probe the region of the parameter space with DP masses of $6-400$ $\mu$eV. Once detected, analysis of signal modulation will allow for a definitive test that DM has been observed and could shed light on the production mechanism.

\acknowledgments
We thank Francesco Capozzi, J\`on Gudmundsson, Matthew Lawson, Georg Raffelt, Karl Van Bibber and Frank Wilczek for helpful discussions. AM thanks Thierry Grenet for pointing out the factor of $2\pi$ error in Ref.~\cite{Lawson:2019brd}. AM is supported by the
European Research Council under Grant~No.~742104 and is supported in part by the research environment grant ``Detecting Axion Dark Matter In The Sky And In The Lab (AxionDM)" funded by the Swedish Research Council (VR) under Dnr 2019-02337.
The work of GG, VT and EV was supported by the U.S. Department of Energy (DOE) Grant No. DE-SC0009937.

\appendix
\section{Absorption rate from thermal field theory}\label{alternative}

Here we employ thermal field theory to obtain the DP DM absorption rate, following the discussion of Refs.~\cite{Caputo:2020quz,Redondo:2013lna}. The emission and absorption rates of a boson by a medium is related to the self-energy of the particle in the medium itself as~\cite{Weldon:1983jn, Kapusta:2006pm}
\begin{equation}
	{\rm Im}\,\Pi=-\omega \Gamma\, ,
	\label{weldon}
\end{equation}
where ${\Gamma=\Gamma_{\rm abs}-\Gamma_{\rm prod}}$ is the rate with which the considered particle distributions approach thermal equilibrium. 
Thus, the absorption rate is obtained by calculating the in-medium DP self-energy. This statement corresponds to the optical theorem in the framework of thermal field theory.

We need to generalize the approach used for dark photon production by an electromagnetic plasma as in Ref.~\cite{Redondo:2013lna} to the case in which a non-thermal population of cold DM is absorbed by a detector. Let us use the momentum distribution $f_X(\bold{k})$ so that the DP number density $n$ is 
 \begin{align}\label{app_numb} 
     n=\int \frac{d^3 \bold{k}}{(2\pi)^3} f_X \, .
 \end{align}
For a boson with absorption rate $\Gamma_{\rm abs}$ and production rate $\Gamma_{\rm prod}$ one finds~\cite{Weldon:1983jn}
\begin{equation}
    \frac{\partial f}{\partial t}=-f\Gamma_{\rm abs}+(1+f)\Gamma_{\rm prod}\, .
\end{equation}
 We consider that only $A$ and $X$ interconvert in the medium so that the production rate of one  is equal to the absorption rate of the other. Since the plasma haloscope is at cryogenic temperatures there is no separate population of photons in the medium. For the  DP masses considered here the occupation number is large, $f_X\gg 1$. Hence, the DP and photon evolve respectively as
\begin{subequations}
\begin{align}
    \frac{\partial f_X}{\partial t}&\simeq-f_X\Gamma^{ X}\\
    \frac{\partial f_A}{\partial t}&=-f_A\Gamma^{ X}_{\rm prod}+(1+f_A)\Gamma^{ X}_{\rm abs}\, ,
\end{align}
\end{subequations}
where $\Gamma^{ X}=\Gamma^{ X}_{\rm abs}-\Gamma^{ X}_{\rm prod}$.
Here, we are assuming that the passage from vacuum to the haloscope is strongly non-adiabatic, so that the components in the basis of interaction eigenstates of the vacuum propagation eigenstate constituting the DM are conserved after the interface. This means that the active component is small right after passing through the medium interface (as it is proportional to $\chi$, not $\chi/\epsilon$), while the sterile component is large. This is analogous to the case of neutrinos oscillating in a matter potential in the ``slab approximation"~\cite{Kuo:1989qe, Giunti:2007ry}.

Using $\partial f_A/\partial t=-\partial f_X/\partial t$, one has
\begin{equation}
    \Gamma^{ X}_{\rm abs}\simeq(f_X-f_A)\Gamma^{ X}\simeq -f_X\frac{{\rm Im}\Pi_{X}}{\omega}\, ,
\end{equation}
where in the last approximation we have assumed that $f_X\gg f_A$  and used Eq.~\eqref{weldon}.

\begin{figure}\vspace{-1.5cm}
  \includegraphics[,width=0.9\linewidth]{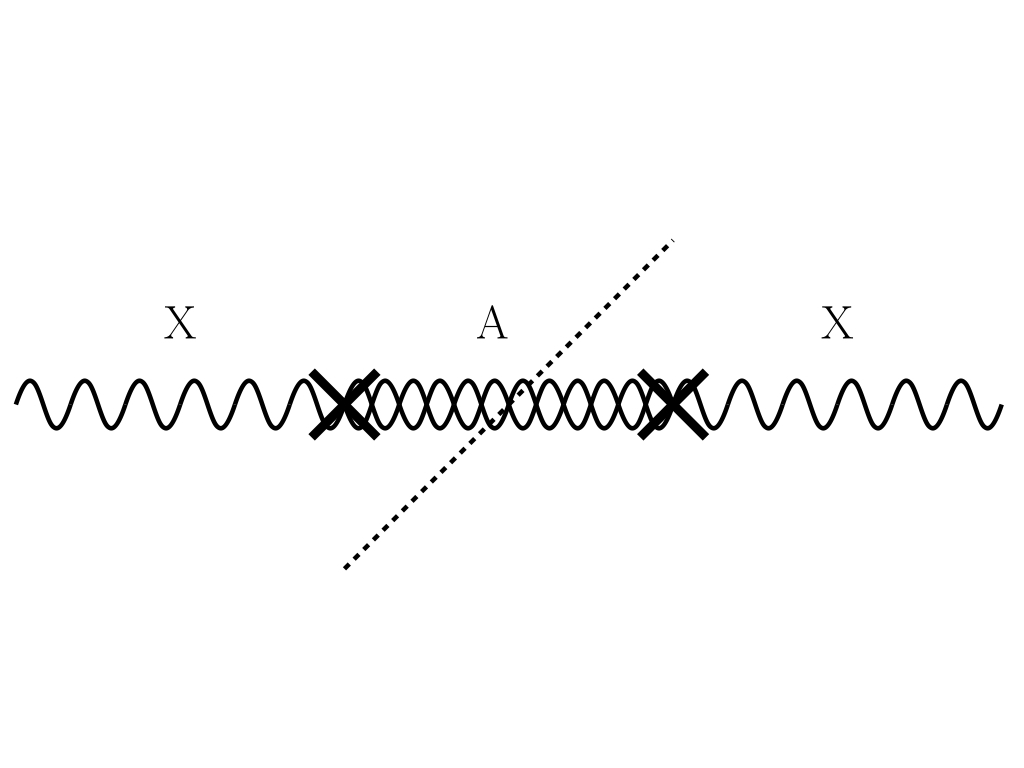}\vspace{-1.2cm}
\caption{Feynman diagram representing the DP (single line) self energy contribution due to the mixing with the plasmon (double line), through the vertex $\chi m_X^2$ (crosses). The plasmon propagator is cut by the diagonal dashed line to indicate that its frequency is taken to be equal to the DP mass as its momentum goes to zero, i.e. $\omega=m_X=\omega_p$.
}
\label{fig:oscill}
\end{figure}

At lowest order the DP self-energy, shown in Fig.~\ref{fig:oscill}, is given by~\cite{Redondo:2013lna}
\begin{align}\label{selfenergy}
	\Pi_{X}^{L,T}=m_X^2+m_X^2 \chi\frac{1}{K^2-\Pi_{L,T}(K)}m_X^2 \chi~,
\end{align}
where $K=(\omega,\bf{k})$ is the four-momentum of the external DP and $\Pi_{L,T}$ are the self-energy of the longitudinal and transverse plasmons respectively (we drop $A$ in the notation). Both longitudinal and transverse excitations can be produced by the DP population, which can be absorbed both when with longitudinal and transverse polarization.

From Eq.~\eqref{selfenergy}, it is seen that we need the plasmon self-energy to obtain the DP absorption rate. The real part, which modifies the dispersion relations of photons in the medium,
 \begin{align}
     \omega^2-k^2=\rm{Re }\, \Pi_{T,L} \, ,
 \end{align}
in the nonrelativistic approximation is given by~\cite{Redondo:2013lna}
 \begin{subequations}
\begin{alignat}{2}
	{\rm Re}\, \Pi_{T}&=\omega_p^2\, ,\\
	{\rm Re}\,\Pi_{L}&=\frac{K^2}{\omega^2}\omega_p^2\, ,
\end{alignat}
\end{subequations}
where $\omega_p$ is the plasma frequency. These are Eqs.~(6.32) and (6.38) of Ref.~\cite{Raffelt:1996wa}, considering that they are written for the case in which Im~$\Pi_{T,L}=0$.
We see that the dispersion relation for the transverse plasmon gives $\omega^2~-~k^2=~\omega_p^2$, with the usual interpretation of transverse excitations as particles with mass $\omega_p$. The latter is given in a system of wires by Eq.~\eqref{eq:pasmafreq}. 
The longitudinal plasmon, on the other hand, has a peculiar dispersion relation, so that in the nonrelativistic limit $\omega$ is independent from $k$.
The imaginary part of the photon self-energy is related to the rate $\Gamma$ by which plasmons thermalize in the medium --- i.e., the damping in the classical description.

When treating longitudinal plasmons one defines the vertex renormalization constant $Z_L$~\cite{An:2013yfc}
\begin{equation}
	K^2=Z_L^{-1}\omega^2\, ,
\end{equation}
relevant for the coupling of external photons
or plasmons to electrons in the medium. We then interpret
\begin{align}
    \Gamma_L=-Z_L{\rm Im}\,\Pi_{L}/\omega \, .
\end{align}
 Enforcing the on-shell condition $K^2=m_X^2$, the resulting absorption rate of DPs is
\begin{align}
	\Gamma^{X}_{\rm abs} &=f_X^L m_X^2\omega^2\frac{ \chi^2 \Gamma_L}{(\omega^2-\omega_p^2)^2+(\omega \Gamma_L)^2}\nonumber\\
	&+f_X^T m_X^4\frac{  \chi^2 \Gamma_T}{(m_X^2-\omega_p^2)^2+(\omega \Gamma_T)^2} \,  ,
\end{align}
where we have defined the longitudinal and transverse DP populations ($f_X^L$ and $f_X^T$, respectively, and $f_X^L+f_X^T=f_X$), and we have used Eq.~\eqref{weldon} to define ${\Gamma_T=-{\rm Im}\,\Pi_T/\omega}$. 
This expression closely resembles the production rate of $X$ bosons obtained in Ref.~\cite{Hardy:2016kme}. Moreover, interpreting
\begin{align}
	\chi_{\rm eff}^2=\chi^2 \frac{m_X^4}{(m_X^2-\omega_p^2)^2+(\omega \Gamma_T)^2}
\end{align}
as an effective coupling, we find the same expression as Eq.~(4.4) of~Ref.~\cite{Hochberg:2017wce} (see also Refs.~\cite{Coskuner:2019odd,An:2014twa}).

The longitudinal and transverse photons are indistinguishable in the zero momentum limit,  ${\Gamma_T=\Gamma_L\equiv \Gamma}$, so that for non-relativistic DPs on resonance ($\omega=m_X=\omega_p$) the number absorption rate per unit volume is
\begin{equation}
	\Gamma^{X}_{\rm abs}=f_X\frac{ \chi^2 m_X^2}{\Gamma}\, .
\end{equation}

As long as the dark matter distribution is narrower than the line width of the resonance, the DP momentum distribution can be written as
\begin{equation}
    f_X=n(2\pi)^3\delta^3({\bf k})\, ,
\end{equation}
where $n= \rho/ m_X$ is the DM number density, which coincides with Eqs.~\eqref{f(p)-with-delta} and \eqref{density-X0}. 
Thus, the power absorbed in a homogeneous and isotropic detector with volume $V_d$ is given by
\begin{equation}
	P=V_d\int \frac{d^3{\bf k}}{(2\pi)^3}\, \omega  \Gamma^{X}_{\rm abs} = \chi^2 \rho \, m_X^2\frac{Q}{m_X}\,V_d \, ,\label{eq:plasmapower}
\end{equation}
	where ${Q=\omega/\Gamma}$ is the quality factor and $\rho$ is the local DM density. Equation~\eqref{eq:plasmapower} is equivalent to Eq.~\eqref{eq:power2} in the main text
	\begin{equation}
P_{\rm out}=\kappa P \, ,
\end{equation}
	in the limit where boundary conditions are negligible so that the ``geometric factor" goes to unity.

\bibliographystyle{bibi}
\bibliography{bibliography}

\providecommand{\href}[2]{#2}\begingroup\raggedright\begin{thebibliography}{100}

\bibitem{Gelmini:2015zpa}
G.~B. Gelmini, \emph{{The Hunt for Dark Matter}},  in \emph{{TASI}: {Journeys
  Through the Precision Frontier: Amplitudes for Colliders}}, pp.~559--616,
  2015, \href{https://arxiv.org/abs/1502.01320}{{\ttfamily 1502.01320}},
  \href{https://doi.org/10.1142/9789814678766\_0012}{DOI}.

\bibitem{Jaeckel:2010ni}
J.~Jaeckel and A.~Ringwald, \emph{{The Low-Energy Frontier of Particle
  Physics}},
  \href{https://doi.org/10.1146/annurev.nucl.012809.104433}{\emph{Ann. Rev.
  Nucl. Part. Sci.} {\bfseries 60} (2010) 405}
  [\href{https://arxiv.org/abs/1002.0329}{{\ttfamily 1002.0329}}].

\bibitem{Pospelov:2008jk}
M.~Pospelov, A.~Ritz and M.~B. Voloshin, \emph{{Bosonic super-WIMPs as
  keV-scale dark matter}},
  \href{https://doi.org/10.1103/PhysRevD.78.115012}{\emph{Phys. Rev. D}
  {\bfseries 78} (2008) 115012}
  [\href{https://arxiv.org/abs/0807.3279}{{\ttfamily 0807.3279}}].

\bibitem{An:2013yua}
H.~An, M.~Pospelov and J.~Pradler, \emph{{Dark Matter Detectors as Dark Photon
  Helioscopes}},
  \href{https://doi.org/10.1103/PhysRevLett.111.041302}{\emph{Phys. Rev. Lett.}
  {\bfseries 111} (2013) 041302}
  [\href{https://arxiv.org/abs/1304.3461}{{\ttfamily 1304.3461}}].

\bibitem{Goodsell:2009xc}
M.~Goodsell, J.~Jaeckel, J.~Redondo and A.~Ringwald, \emph{{Naturally Light
  Hidden Photons in LARGE Volume String Compactifications}},
  \href{https://doi.org/10.1088/1126-6708/2009/11/027}{\emph{JHEP} {\bfseries
  11} (2009) 027} [\href{https://arxiv.org/abs/0909.0515}{{\ttfamily
  0909.0515}}].

\bibitem{Preskill:1982cy}
J.~Preskill, M.~B. Wise and F.~Wilczek, \emph{{Cosmology of the Invisible
  Axion}}, \href{https://doi.org/10.1016/0370-2693(83)90637-8}{\emph{Phys.
  Lett. B} {\bfseries 120} (1983) 127}.

\bibitem{Abbott:1982af}
L.~Abbott and P.~Sikivie, \emph{{A Cosmological Bound on the Invisible Axion}},
  \href{https://doi.org/10.1016/0370-2693(83)90638-X}{\emph{Phys. Lett. B}
  {\bfseries 120} (1983) 133}.

\bibitem{Dine:1982ah}
M.~Dine and W.~Fischler, \emph{{The Not So Harmless Axion}},
  \href{https://doi.org/10.1016/0370-2693(83)90639-1}{\emph{Phys. Lett. B}
  {\bfseries 120} (1983) 137}.

\bibitem{Nelson:2011sf}
A.~E. Nelson and J.~Scholtz, \emph{{Dark Light, Dark Matter and the
  Misalignment Mechanism}},
  \href{https://doi.org/10.1103/PhysRevD.84.103501}{\emph{Phys. Rev. D}
  {\bfseries 84} (2011) 103501}
  [\href{https://arxiv.org/abs/1105.2812}{{\ttfamily 1105.2812}}].

\bibitem{Arias:2012az}
P.~Arias, D.~Cadamuro, M.~Goodsell, J.~Jaeckel, J.~Redondo and A.~Ringwald,
  \emph{{WISPy Cold Dark Matter}},
  \href{https://doi.org/10.1088/1475-7516/2012/06/013}{\emph{JCAP} {\bfseries
  06} (2012) 013} [\href{https://arxiv.org/abs/1201.5902}{{\ttfamily
  1201.5902}}].

\bibitem{AlonsoAlvarez:2019cgw}
G.~Alonso-Álvarez, T.~Hugle and J.~Jaeckel, \emph{{Misalignment \& Co.:
  (Pseudo-)scalar and vector dark matter with curvature couplings}},
  \href{https://doi.org/10.1088/1475-7516/2020/02/014}{\emph{JCAP} {\bfseries
  02} (2020) 014} [\href{https://arxiv.org/abs/1905.09836}{{\ttfamily
  1905.09836}}].

\bibitem{Nakayama:2019rhg}
K.~Nakayama, \emph{{Vector Coherent Oscillation Dark Matter}},
  \href{https://doi.org/10.1088/1475-7516/2019/10/019}{\emph{JCAP} {\bfseries
  10} (2019) 019} [\href{https://arxiv.org/abs/1907.06243}{{\ttfamily
  1907.06243}}].

\bibitem{Nakayama:2020rka}
K.~Nakayama, \emph{{Constraint on Vector Coherent Oscillation Dark Matter with
  Kinetic Function}},  \href{https://arxiv.org/abs/2004.10036}{{\ttfamily
  2004.10036}}.

\bibitem{Graham:2015rva}
P.~W. Graham, J.~Mardon and S.~Rajendran, \emph{{Vector Dark Matter from
  Inflationary Fluctuations}},
  \href{https://doi.org/10.1103/PhysRevD.93.103520}{\emph{Phys. Rev. D}
  {\bfseries 93} (2016) 103520}
  [\href{https://arxiv.org/abs/1504.02102}{{\ttfamily 1504.02102}}].

\bibitem{Nakai:2020cfw}
Y.~Nakai, R.~Namba and Z.~Wang, \emph{{Light Dark Photon Dark Matter from
  Inflation}},  \href{https://arxiv.org/abs/2004.10743}{{\ttfamily
  2004.10743}}.

\bibitem{Dror:2018pdh}
J.~A. Dror, K.~Harigaya and V.~Narayan, \emph{{Parametric Resonance Production
  of Ultralight Vector Dark Matter}},
  \href{https://doi.org/10.1103/PhysRevD.99.035036}{\emph{Phys. Rev. D}
  {\bfseries 99} (2019) 035036}
  [\href{https://arxiv.org/abs/1810.07195}{{\ttfamily 1810.07195}}].

\bibitem{Agrawal:2018vin}
P.~Agrawal, N.~Kitajima, M.~Reece, T.~Sekiguchi and F.~Takahashi, \emph{{Relic
  Abundance of Dark Photon Dark Matter}},
  \href{https://doi.org/10.1016/j.physletb.2019.135136}{\emph{Phys. Lett. B}
  {\bfseries 801} (2020) 135136}
  [\href{https://arxiv.org/abs/1810.07188}{{\ttfamily 1810.07188}}].

\bibitem{Co:2018lka}
R.~T. Co, A.~Pierce, Z.~Zhang and Y.~Zhao, \emph{{Dark Photon Dark Matter
  Produced by Axion Oscillations}},
  \href{https://doi.org/10.1103/PhysRevD.99.075002}{\emph{Phys. Rev. D}
  {\bfseries 99} (2019) 075002}
  [\href{https://arxiv.org/abs/1810.07196}{{\ttfamily 1810.07196}}].

\bibitem{Bastero-Gil:2018uel}
M.~Bastero-Gil, J.~Santiago, L.~Ubaldi and R.~Vega-Morales, \emph{{Vector dark
  matter production at the end of inflation}},
  \href{https://doi.org/10.1088/1475-7516/2019/04/015}{\emph{JCAP} {\bfseries
  04} (2019) 015} [\href{https://arxiv.org/abs/1810.07208}{{\ttfamily
  1810.07208}}].

\bibitem{Long:2019lwl}
A.~J. Long and L.-T. Wang, \emph{{Dark Photon Dark Matter from a Network of
  Cosmic Strings}},
  \href{https://doi.org/10.1103/PhysRevD.99.063529}{\emph{Phys. Rev. D}
  {\bfseries 99} (2019) 063529}
  [\href{https://arxiv.org/abs/1901.03312}{{\ttfamily 1901.03312}}].

\bibitem{Aprile:2017iyp}
{\scshape XENON} Collaboration, E.~Aprile et~al., \emph{{First Dark Matter
  Search Results from the XENON1T Experiment}},
  \href{https://doi.org/10.1103/PhysRevLett.119.181301}{\emph{Phys. Rev. Lett.}
  {\bfseries 119} (2017) 181301}
  [\href{https://arxiv.org/abs/1705.06655}{{\ttfamily 1705.06655}}].

\bibitem{Aprile:2018dbl}
{\scshape XENON} Collaboration, E.~Aprile et~al., \emph{{Dark Matter Search
  Results from a One Ton-Year Exposure of XENON1T}},
  \href{https://doi.org/10.1103/PhysRevLett.121.111302}{\emph{Phys. Rev. Lett.}
  {\bfseries 121} (2018) 111302}
  [\href{https://arxiv.org/abs/1805.12562}{{\ttfamily 1805.12562}}].

\bibitem{Akerib:2016vxi}
{\scshape LUX} Collaboration, D.~Akerib et~al., \emph{{Results from a search
  for dark matter in the complete LUX exposure}},
  \href{https://doi.org/10.1103/PhysRevLett.118.021303}{\emph{Phys. Rev. Lett.}
  {\bfseries 118} (2017) 021303}
  [\href{https://arxiv.org/abs/1608.07648}{{\ttfamily 1608.07648}}].

\bibitem{An:2014twa}
H.~An, M.~Pospelov, J.~Pradler and A.~Ritz, \emph{{Direct Detection Constraints
  on Dark Photon Dark Matter}},
  \href{https://doi.org/10.1016/j.physletb.2015.06.018}{\emph{Phys. Lett. B}
  {\bfseries 747} (2015) 331}
  [\href{https://arxiv.org/abs/1412.8378}{{\ttfamily 1412.8378}}].

\bibitem{Bloch:2016sjj}
I.~M. Bloch, R.~Essig, K.~Tobioka, T.~Volansky and T.-T. Yu, \emph{{Searching
  for Dark Absorption with Direct Detection Experiments}},
  \href{https://doi.org/10.1007/JHEP06(2017)087}{\emph{JHEP} {\bfseries 06}
  (2017) 087} [\href{https://arxiv.org/abs/1608.02123}{{\ttfamily
  1608.02123}}].

\bibitem{Hochberg:2015pha}
Y.~Hochberg, Y.~Zhao and K.~M. Zurek, \emph{{Superconducting Detectors for
  Superlight Dark Matter}},
  \href{https://doi.org/10.1103/PhysRevLett.116.011301}{\emph{Phys. Rev. Lett.}
  {\bfseries 116} (2016) 011301}
  [\href{https://arxiv.org/abs/1504.07237}{{\ttfamily 1504.07237}}].

\bibitem{Hochberg:2015fth}
Y.~Hochberg, M.~Pyle, Y.~Zhao and K.~M. Zurek, \emph{{Detecting Superlight Dark
  Matter with Fermi-Degenerate Materials}},
  \href{https://doi.org/10.1007/JHEP08(2016)057}{\emph{JHEP} {\bfseries 08}
  (2016) 057} [\href{https://arxiv.org/abs/1512.04533}{{\ttfamily
  1512.04533}}].

\bibitem{Hochberg:2016ntt}
Y.~Hochberg, Y.~Kahn, M.~Lisanti, C.~G. Tully and K.~M. Zurek,
  \emph{{Directional detection of dark matter with two-dimensional targets}},
  \href{https://doi.org/10.1016/j.physletb.2017.06.051}{\emph{Phys. Lett. B}
  {\bfseries 772} (2017) 239}
  [\href{https://arxiv.org/abs/1606.08849}{{\ttfamily 1606.08849}}].

\bibitem{Hochberg:2017wce}
Y.~Hochberg, Y.~Kahn, M.~Lisanti, K.~M. Zurek, A.~G. Grushin, R.~Ilan, S.~M.
  Griffin, Z.-F. Liu, S.~F. Weber and J.~B. Neaton, \emph{{Detection of sub-MeV
  Dark Matter with Three-Dimensional Dirac Materials}},
  \href{https://doi.org/10.1103/PhysRevD.97.015004}{\emph{Phys. Rev. D}
  {\bfseries 97} (2018) 015004}
  [\href{https://arxiv.org/abs/1708.08929}{{\ttfamily 1708.08929}}].

\bibitem{Coskuner:2019odd}
A.~Coskuner, A.~Mitridate, A.~Olivares and K.~M. Zurek, \emph{{Directional Dark
  Matter Detection in Anisotropic Dirac Materials}},
  \href{https://arxiv.org/abs/1909.09170}{{\ttfamily 1909.09170}}.

\bibitem{Knapen:2016cue}
S.~Knapen, T.~Lin and K.~M. Zurek, \emph{{Light Dark Matter in Superfluid
  Helium: Detection with Multi-excitation Production}},
  \href{https://doi.org/10.1103/PhysRevD.95.056019}{\emph{Phys. Rev. D}
  {\bfseries 95} (2017) 056019}
  [\href{https://arxiv.org/abs/1611.06228}{{\ttfamily 1611.06228}}].

\bibitem{Schutz:2016tid}
K.~Schutz and K.~M. Zurek, \emph{{Detectability of Light Dark Matter with
  Superfluid Helium}},
  \href{https://doi.org/10.1103/PhysRevLett.117.121302}{\emph{Phys. Rev. Lett.}
  {\bfseries 117} (2016) 121302}
  [\href{https://arxiv.org/abs/1604.08206}{{\ttfamily 1604.08206}}].

\bibitem{Caputo:2019cyg}
A.~Caputo, A.~Esposito and A.~D. Polosa, \emph{{Sub-MeV Dark Matter and the
  Goldstone Modes of Superfluid Helium}},
  \href{https://doi.org/10.1103/PhysRevD.100.116007}{\emph{Phys. Rev. D}
  {\bfseries 100} (2019) 116007}
  [\href{https://arxiv.org/abs/1907.10635}{{\ttfamily 1907.10635}}].

\bibitem{Acanfora:2019con}
F.~Acanfora, A.~Esposito and A.~D. Polosa, \emph{{Sub-GeV Dark Matter in
  Superfluid He-4: an Effective Theory Approach}},
  \href{https://doi.org/10.1140/epjc/s10052-019-7057-0}{\emph{Eur. Phys. J. C}
  {\bfseries 79} (2019) 549}
  [\href{https://arxiv.org/abs/1902.02361}{{\ttfamily 1902.02361}}].

\bibitem{Guo:2013dt}
W.~Guo and D.~N. McKinsey, \emph{{Concept for a dark matter detector using
  liquid helium-4}},
  \href{https://doi.org/10.1103/PhysRevD.87.115001}{\emph{Phys. Rev. D}
  {\bfseries 87} (2013) 115001}
  [\href{https://arxiv.org/abs/1302.0534}{{\ttfamily 1302.0534}}].

\bibitem{Knapen:2017ekk}
S.~Knapen, T.~Lin, M.~Pyle and K.~M. Zurek, \emph{{Detection of Light Dark
  Matter With Optical Phonons in Polar Materials}},
  \href{https://doi.org/10.1016/j.physletb.2018.08.064}{\emph{Phys. Lett. B}
  {\bfseries 785} (2018) 386}
  [\href{https://arxiv.org/abs/1712.06598}{{\ttfamily 1712.06598}}].

\bibitem{Irastorza:2018dyq}
I.~G. Irastorza and J.~Redondo, \emph{{New experimental approaches in the
  search for axion-like particles}},
  \href{https://doi.org/10.1016/j.ppnp.2018.05.003}{\emph{Prog. Part. Nucl.
  Phys.} {\bfseries 102} (2018) 89}
  [\href{https://arxiv.org/abs/1801.08127}{{\ttfamily 1801.08127}}].

\bibitem{Sikivie:1983ip}
P.~Sikivie, \emph{{Experimental Tests of the Invisible Axion}},
  \href{https://doi.org/10.1103/PhysRevLett.51.1415}{\emph{Phys. Rev. Lett.}
  {\bfseries 51} (1983) 1415}. [Erratum: Phys.Rev.Lett. 52, 695 (1984)].

\bibitem{Rybka:2014xca}
{\scshape ADMX} Collaboration, G.~Rybka, \emph{{Direct detection searches for
  axion dark matter}},  in \emph{{Proceedings, 13th international conference on
  Topics in Astroparticle and Underground Physics (TAUP 2013): Asilomar,
  California, September 8-13, 2013}},
  \href{https://doi.org/10.1016/j.dark.2014.05.003}{DOI}.

\bibitem{Woohyun:2016}
W.~Chung, \emph{{Launching axion experiment at CAPP/IBS in Korea}},  in
  \emph{{Proceedings, 12th Patras workshop on axions, WIMPs and WISPs: Jeju
  Island, South Korea, June 20-24, 2016}}, DESY: Hamburg, Germany (2017)
  30--34, \href{https://doi.org/10.3204/DESY-PROC-2009-03/Chung_Woohyun}{DOI}.

\bibitem{Goryachev:2017wpw}
M.~Goryachev, B.~T. Mcallister and M.~E. Tobar, \emph{{Axion detection with
  negatively coupled cavity arrays}},
  \href{https://doi.org/10.1016/j.physleta.2017.09.016}{\emph{Phys. Lett. A}
  {\bfseries 382} (2018) 2199}
  [\href{https://arxiv.org/abs/1703.07207}{{\ttfamily 1703.07207}}].

\bibitem{Alesini:2017ifp}
D.~Alesini, D.~Babusci, D.~Di~Gioacchino, C.~Gatti, G.~Lamanna and C.~Ligi,
  \emph{{The KLASH Proposal}},
  \href{https://arxiv.org/abs/1707.06010}{{\ttfamily 1707.06010}}.

\bibitem{Melcon:2018dba}
A.~A. Melc\'on et~al., \emph{{Axion Searches with Microwave Filters: the RADES
  project}}, \href{https://doi.org/10.1088/1475-7516/2018/05/040}{\emph{JCAP}
  {\bfseries 05} (2018) 040}
  [\href{https://arxiv.org/abs/1803.01243}{{\ttfamily 1803.01243}}].

\bibitem{Melcon:2020xvj}
A.~A. Melc\'on et~al., \emph{{Scalable haloscopes for axion dark matter
  detection in the 30$\mu$eV range with RADES}},
  \href{https://arxiv.org/abs/2002.07639}{{\ttfamily 2002.07639}}.

\bibitem{TheMADMAXWorkingGroup:2016hpc}
{\scshape MADMAX Working Group} Collaboration, A.~Caldwell, G.~Dvali,
  B.~Majorovits, A.~Millar, G.~Raffelt, J.~Redondo, O.~Reimann, F.~Simon and
  F.~Steffen, \emph{{Dielectric Haloscopes: A New Way to Detect Axion Dark
  Matter}}, \href{https://doi.org/10.1103/PhysRevLett.118.091801}{\emph{Phys.
  Rev. Lett.} {\bfseries 118} (2017) 091801}
  [\href{https://arxiv.org/abs/1611.05865}{{\ttfamily 1611.05865}}].

\bibitem{Baryakhtar:2018doz}
M.~Baryakhtar, J.~Huang and R.~Lasenby, \emph{{Axion and hidden photon dark
  matter detection with multilayer optical haloscopes}},
  \href{https://doi.org/10.1103/PhysRevD.98.035006}{\emph{Phys. Rev. D}
  {\bfseries 98} (2018) 035006}
  [\href{https://arxiv.org/abs/1803.11455}{{\ttfamily 1803.11455}}].

\bibitem{Jaeckel:2013sqa}
J.~Jaeckel and J.~Redondo, \emph{{An antenna for directional detection of WISPy
  dark matter}},
  \href{https://doi.org/10.1088/1475-7516/2013/11/016}{\emph{JCAP} {\bfseries
  11} (2013) 016} [\href{https://arxiv.org/abs/1307.7181}{{\ttfamily
  1307.7181}}].

\bibitem{Horns:2012jf}
D.~Horns, J.~Jaeckel, A.~Lindner, A.~Lobanov, J.~Redondo and A.~Ringwald,
  \emph{{Searching for WISPy Cold Dark Matter with a Dish Antenna}},
  \href{https://doi.org/10.1088/1475-7516/2013/04/016}{\emph{JCAP} {\bfseries
  04} (2013) 016} [\href{https://arxiv.org/abs/1212.2970}{{\ttfamily
  1212.2970}}].

\bibitem{Suzuki:2015sza}
J.~Suzuki, T.~Horie, Y.~Inoue and M.~Minowa, \emph{{Experimental Search for
  Hidden Photon CDM in the eV mass range with a Dish Antenna}},
  \href{https://doi.org/10.1088/1475-7516/2015/09/042}{\emph{JCAP} {\bfseries
  09} (2015) 042} [\href{https://arxiv.org/abs/1504.00118}{{\ttfamily
  1504.00118}}].

\bibitem{Experiment:2017icw}
{\scshape FUNK Experiment} Collaboration, D.~Veberi\v~c et~al., \emph{{Search
  for hidden-photon dark matter with the FUNK experiment}},
  \href{https://doi.org/10.22323/1.301.0880}{\emph{PoS} {\bfseries ICRC2017}
  (2018) 880} [\href{https://arxiv.org/abs/1711.02958}{{\ttfamily
  1711.02958}}].

\bibitem{BRASS}
\url{http://www.iexp.uni-hamburg.de/groups/astroparticle/brass/brassweb.htm}.

\bibitem{Lawson:2019brd}
M.~Lawson, A.~J. Millar, M.~Pancaldi, E.~Vitagliano and F.~Wilczek,
  \emph{{Tunable axion plasma haloscopes}},
  \href{https://doi.org/10.1103/PhysRevLett.123.141802}{\emph{Phys. Rev. Lett.}
  {\bfseries 123} (2019) 141802}
  [\href{https://arxiv.org/abs/1904.11872}{{\ttfamily 1904.11872}}].

\bibitem{Raffelt:1996wa}
G.~G. Raffelt, \emph{{Stars as laboratories for fundamental physics}}. Chicago,
  USA: Univ. Pr., 1996.

\bibitem{An:2013yfc}
H.~An, M.~Pospelov and J.~Pradler, \emph{{New stellar constraints on dark
  photons}}, \href{https://doi.org/10.1016/j.physletb.2013.07.008}{\emph{Phys.
  Lett. B} {\bfseries 725} (2013) 190}
  [\href{https://arxiv.org/abs/1302.3884}{{\ttfamily 1302.3884}}].

\bibitem{Redondo:2013lna}
J.~Redondo and G.~Raffelt, \emph{{Solar constraints on hidden photons
  re-visited}},
  \href{https://doi.org/10.1088/1475-7516/2013/08/034}{\emph{JCAP} {\bfseries
  08} (2013) 034} [\href{https://arxiv.org/abs/1305.2920}{{\ttfamily
  1305.2920}}].

\bibitem{Dvorkin:2019zdi}
C.~Dvorkin, T.~Lin and K.~Schutz, \emph{{Making dark matter out of light:
  freeze-in from plasma effects}},
  \href{https://doi.org/10.1103/PhysRevD.99.115009}{\emph{Phys. Rev. D}
  {\bfseries 99} (2019) 115009}
  [\href{https://arxiv.org/abs/1902.08623}{{\ttfamily 1902.08623}}].

\bibitem{Kurinsky:2020dpb}
N.~Kurinsky, D.~Baxter, Y.~Kahn and G.~Krnjaic, \emph{{A Dark Matter
  Interpretation of Excesses in Multiple Direct Detection Experiments}},
  \href{https://arxiv.org/abs/2002.06937}{{\ttfamily 2002.06937}}.

\bibitem{Kozaczuk:2020uzb}
J.~Kozaczuk and T.~Lin, \emph{{Plasmon production from dark matter
  scattering}}, \href{https://doi.org/10.1103/PhysRevD.101.123012}{\emph{Phys.
  Rev. D} {\bfseries 101} (2020) 123012}
  [\href{https://arxiv.org/abs/2003.12077}{{\ttfamily 2003.12077}}].

\bibitem{Holdom:1985ag}
B.~Holdom, \emph{{Two U(1)'s and Epsilon Charge Shifts}},
  \href{https://doi.org/10.1016/0370-2693(86)91377-8}{\emph{Phys. Lett. B}
  {\bfseries 166} (1986) 196}.

\bibitem{Fabbrichesi:2020wbt}
M.~Fabbrichesi, E.~Gabrielli and G.~Lanfranchi, \emph{{The Dark Photon}},
  \href{https://arxiv.org/abs/2005.01515}{{\ttfamily 2005.01515}}.

\bibitem{Knirck:2018knd}
S.~Knirck, A.~J. Millar, C.~A. O'Hare, J.~Redondo and F.~D. Steffen,
  \emph{{Directional axion detection}},
  \href{https://doi.org/10.1088/1475-7516/2018/11/051}{\emph{JCAP} {\bfseries
  11} (2018) 051} [\href{https://arxiv.org/abs/1806.05927}{{\ttfamily
  1806.05927}}].

\bibitem{Haft:1993jt}
M.~Haft, G.~Raffelt and A.~Weiss, \emph{{Standard and nonstandard plasma
  neutrino emission revisited}},
  \href{https://doi.org/10.1086/173978}{\emph{Astrophys. J.} {\bfseries 425}
  (1994) 222} [\href{https://arxiv.org/abs/astro-ph/9309014}{{\ttfamily
  astro-ph/9309014}}]. [Erratum: Astrophys.J. 438, 1017 (1995)].

\bibitem{Weldon:1982aq}
H.~A. Weldon, \emph{{Covariant Calculations at Finite Temperature: The
  Relativistic Plasma}},
  \href{https://doi.org/10.1103/PhysRevD.26.1394}{\emph{Phys. Rev. D}
  {\bfseries 26} (1982) 1394}.

\bibitem{McDermott:2019lch}
S.~D. McDermott and S.~J. Witte, \emph{{Cosmological evolution of light dark
  photon dark matter}},
  \href{https://doi.org/10.1103/PhysRevD.101.063030}{\emph{Phys. Rev. D}
  {\bfseries 101} (2020) 063030}
  [\href{https://arxiv.org/abs/1911.05086}{{\ttfamily 1911.05086}}].

\bibitem{Witte:2020rvb}
S.~J. Witte, S.~Rosauro-Alcaraz, S.~D. McDermott and V.~Poulin, \emph{{Dark
  photon dark matter in the presence of inhomogeneous structure}},
  \href{https://doi.org/10.1007/JHEP06(2020)132}{\emph{JHEP} {\bfseries 06}
  (2020) 132} [\href{https://arxiv.org/abs/2003.13698}{{\ttfamily
  2003.13698}}].

\bibitem{Caputo:2020bdy}
A.~Caputo, H.~Liu, S.~Mishra-Sharma and J.~T. Ruderman, \emph{{Dark Photon
  Oscillations in Our Inhomogeneous Universe}},
  \href{https://arxiv.org/abs/2002.05165}{{\ttfamily 2002.05165}}.

\bibitem{Dubovsky:2015cca}
S.~Dubovsky and G.~Hernández-Chifflet, \emph{{Heating up the Galaxy with
  Hidden Photons}},
  \href{https://doi.org/10.1088/1475-7516/2015/12/054}{\emph{JCAP} {\bfseries
  12} (2015) 054} [\href{https://arxiv.org/abs/1509.00039}{{\ttfamily
  1509.00039}}].

\bibitem{Jaeckel:2015kea}
J.~Jaeckel and S.~Knirck, \emph{{Directional Resolution of Dish Antenna
  Experiments to Search for WISPy Dark Matter}},
  \href{https://doi.org/10.1088/1475-7516/2016/01/005}{\emph{JCAP} {\bfseries
  01} (2016) 005} [\href{https://arxiv.org/abs/1509.00371}{{\ttfamily
  1509.00371}}].

\bibitem{landau2013electrodynamics}
L.~D. Landau, J.~Bell, M.~Kearsley, L.~Pitaevskii, E.~Lifshitz and J.~Sykes,
  \emph{Electrodynamics of continuous media}, vol.~8. elsevier, 2013.

\bibitem{Asztalos:2009yp}
{\scshape ADMX} Collaboration, S.~Asztalos et~al., \emph{{A SQUID-based
  microwave cavity search for dark-matter axions}},
  \href{https://doi.org/10.1103/PhysRevLett.104.041301}{\emph{Phys. Rev. Lett.}
  {\bfseries 104} (2010) 041301}
  [\href{https://arxiv.org/abs/0910.5914}{{\ttfamily 0910.5914}}].

\bibitem{deSalas:2019rdi}
P.~F. de~Salas, \emph{{Dark matter local density determination based on recent
  observations}},
  \href{https://doi.org/10.1088/1742-6596/1468/1/012020}{\emph{J. Phys. Conf.
  Ser.} {\bfseries 1468} (2020) 012020}
  [\href{https://arxiv.org/abs/1910.14366}{{\ttfamily 1910.14366}}].

\bibitem{Benito:2019ngh}
M.~Benito, A.~Cuoco and F.~Iocco, \emph{{Handling the Uncertainties in the
  Galactic Dark Matter Distribution for Particle Dark Matter Searches}},
  \href{https://doi.org/10.1088/1475-7516/2019/03/033}{\emph{JCAP} {\bfseries
  03} (2019) 033} [\href{https://arxiv.org/abs/1901.02460}{{\ttfamily
  1901.02460}}].

\bibitem{Read:2014qva}
J.~Read, \emph{{The Local Dark Matter Density}},
  \href{https://doi.org/10.1088/0954-3899/41/6/063101}{\emph{J. Phys. G}
  {\bfseries 41} (2014) 063101}
  [\href{https://arxiv.org/abs/1404.1938}{{\ttfamily 1404.1938}}].

\bibitem{Pendry:1998}
J.~B. Pendry, A.~J. Holden, D.~J. Robbins and W.~J. Stewart, \emph{Low
  frequency plasmons in thin-wire structures},
  \href{https://doi.org/10.1088/0953-8984/10/22/007}{\emph{J. Phys. Condens.
  Matter} {\bfseries 10} (1998) 4785}.

\bibitem{Belov:2003}
P.~A. Belov, R.~Marqués, S.~I. Maslovski, I.~S. Nefedov, M.~Silveirinha, C.~R.
  Simovski and S.~A. Tretyakov, \emph{Strong spatial dispersion in wire media
  in the very large wavelength limit},
  \href{https://doi.org/10.1103/physrevb.67.113103}{\emph{Phys. Rev. B}
  {\bfseries 67} (2003) 113103}.

\bibitem{Stern:2015kzo}
I.~Stern, A.~Chisholm, J.~Hoskins, P.~Sikivie, N.~Sullivan, D.~Tanner,
  G.~Carosi and K.~van Bibber, \emph{{Cavity design for high-frequency axion
  dark matter detectors}}, \href{https://doi.org/10.1063/1.4938164}{\emph{Rev.
  Sci. Instrum.} {\bfseries 86} (2015) 123305}
  [\href{https://arxiv.org/abs/1603.06990}{{\ttfamily 1603.06990}}].

\bibitem{Zhong:2018rsr}
{\scshape HAYSTAC} Collaboration, L.~Zhong et~al., \emph{{Results from phase 1
  of the HAYSTAC microwave cavity axion experiment}},
  \href{https://doi.org/10.1103/PhysRevD.97.092001}{\emph{Phys. Rev. D}
  {\bfseries 97} (2018) 092001}
  [\href{https://arxiv.org/abs/1803.03690}{{\ttfamily 1803.03690}}].

\bibitem{Braine:2019fqb}
{\scshape ADMX} Collaboration, T.~Braine et~al., \emph{{Extended Search for the
  Invisible Axion with the Axion Dark Matter Experiment}},
  \href{https://doi.org/10.1103/PhysRevLett.124.101303}{\emph{Phys. Rev. Lett.}
  {\bfseries 124} (2020) 101303}
  [\href{https://arxiv.org/abs/1910.08638}{{\ttfamily 1910.08638}}].

\bibitem{doi:10.1063/1.1513663}
P.~Gay-Balmaz, C.~Maccio and O.~J.~F. Martin, \emph{Microwire arrays with
  plasmonic response at microwave frequencies},
  \href{https://doi.org/10.1063/1.1513663}{\emph{Applied Physics Letters}
  {\bfseries 81} (2002) 2896}
  [\href{https://arxiv.org/abs/https://doi.org/10.1063/1.1513663}{{\ttfamily
  https://doi.org/10.1063/1.1513663}}].

\bibitem{Ouellet:2018nfr}
J.~Ouellet and Z.~Bogorad, \emph{{Solutions to Axion Electrodynamics in Various
  Geometries}}, \href{https://doi.org/10.1103/PhysRevD.99.055010}{\emph{Phys.
  Rev. D} {\bfseries 99} (2019) 055010}
  [\href{https://arxiv.org/abs/1809.10709}{{\ttfamily 1809.10709}}].

\bibitem{doi:10.1063/1.5126963}
M.~Trepanier, D.~Zhang, L.~V. Filippenko, V.~P. Koshelets and S.~M. Anlage,
  \emph{Tunable superconducting josephson dielectric metamaterial},
  \href{https://doi.org/10.1063/1.5126963}{\emph{AIP Advances} {\bfseries 9}
  (2019) 105320}
  [\href{https://arxiv.org/abs/https://doi.org/10.1063/1.5126963}{{\ttfamily
  https://doi.org/10.1063/1.5126963}}].

\bibitem{Kuo:2019cps}
C.-L. Kuo, \emph{{Large-Volume Centimeter-Wave Cavities for Axion Searches}},
  \href{https://doi.org/10.1088/1475-7516/2020/06/010}{\emph{JCAP} {\bfseries
  06} (2020) 010} [\href{https://arxiv.org/abs/1910.04156}{{\ttfamily
  1910.04156}}].

\bibitem{Chaudhuri:2018rqn}
S.~Chaudhuri, K.~Irwin, P.~W. Graham and J.~Mardon, \emph{{Fundamental Limits
  of Electromagnetic Axion and Hidden-Photon Dark Matter Searches: Part I - The
  Quantum Limit}},  \href{https://arxiv.org/abs/1803.01627}{{\ttfamily
  1803.01627}}.

\bibitem{Millar:2016cjp}
A.~J. Millar, G.~G. Raffelt, J.~Redondo and F.~D. Steffen, \emph{{Dielectric
  Haloscopes to Search for Axion Dark Matter: Theoretical Foundations}},
  \href{https://doi.org/10.1088/1475-7516/2017/01/061}{\emph{JCAP} {\bfseries
  01} (2017) 061} [\href{https://arxiv.org/abs/1612.07057}{{\ttfamily
  1612.07057}}].

\bibitem{Brun:2019kak}
P.~Brun, L.~Chevalier and C.~Flouzat, \emph{{Direct Searches for Hidden-Photon
  Dark Matter with the SHUKET Experiment}},
  \href{https://doi.org/10.1103/PhysRevLett.122.201801}{\emph{Phys. Rev. Lett.}
  {\bfseries 122} (2019) 201801}
  [\href{https://arxiv.org/abs/1905.05579}{{\ttfamily 1905.05579}}].

\bibitem{Suzuki:2015vka}
J.~Suzuki, Y.~Inoue, T.~Horie and M.~Minowa, \emph{{Hidden photon CDM search at
  Tokyo}},  in \emph{{11th Patras Workshop on Axions, WIMPs and WISPs}},
  pp.~145--148, 2015, \href{https://arxiv.org/abs/1509.00785}{{\ttfamily
  1509.00785}},
  \href{https://doi.org/10.3204/DESY-PROC-2015-02/junya\_suzuki}{DOI}.

\bibitem{Wuensch:1989sa}
W.~Wuensch, S.~De~Panfilis-Wuensch, Y.~Semertzidis, J.~Rogers, A.~Melissinos,
  H.~Halama, B.~Moskowitz, A.~Prodell, W.~Fowler and F.~Nezrick, \emph{{Results
  of a Laboratory Search for Cosmic Axions and Other Weakly Coupled Light
  Particles}}, \href{https://doi.org/10.1103/PhysRevD.40.3153}{\emph{Phys. Rev.
  D} {\bfseries 40} (1989) 3153}.

\bibitem{DePanfilis:1987dk}
S.~De~Panfilis, A.~Melissinos, B.~Moskowitz, J.~Rogers, Y.~Semertzidis,
  W.~Wuensch, H.~Halama, A.~Prodell, W.~Fowler and F.~Nezrick, \emph{{Limits on
  the Abundance and Coupling of Cosmic Axions at $4.5~\mu{\rm eV} < m_a <
  5.0~\mu{\rm eV}$}},
  \href{https://doi.org/10.1103/PhysRevLett.59.839}{\emph{Phys. Rev. Lett.}
  {\bfseries 59} (1987) 839}.

\bibitem{Brubaker:2016ktl}
B.~Brubaker et~al., \emph{{First results from a microwave cavity axion search
  at 24 $\mu$eV}},
  \href{https://doi.org/10.1103/PhysRevLett.118.061302}{\emph{Phys. Rev. Lett.}
  {\bfseries 118} (2017) 061302}
  [\href{https://arxiv.org/abs/1610.02580}{{\ttfamily 1610.02580}}].

\bibitem{Brun:2019lyf}
{\scshape MADMAX} Collaboration, P.~Brun et~al., \emph{{A new experimental
  approach to probe QCD axion dark matter in the mass range above 40 $\mu$eV}},
  \href{https://doi.org/10.1140/epjc/s10052-019-6683-x}{\emph{Eur. Phys. J. C}
  {\bfseries 79} (2019) 186}
  [\href{https://arxiv.org/abs/1901.07401}{{\ttfamily 1901.07401}}].

\bibitem{DeMiguel-Hernandez:2020mon}
J.~De~Miguel-Hernández, \emph{{A dark matter telescope probing the 6 to 60 GHz
  band}},  \href{https://arxiv.org/abs/2003.06874}{{\ttfamily 2003.06874}}.

\bibitem{Droster:2019fur}
{\scshape HAYSTAC} Collaboration, A.~Droster and K.~van Bibber, \emph{{HAYSTAC
  Status, Results, and Plans}},  in \emph{{13th Conference on the Intersections
  of Particle and Nuclear Physics}}, 1, 2019,
  \href{https://arxiv.org/abs/1901.01668}{{\ttfamily 1901.01668}}.

\bibitem{Asztalos:2001jk}
{\scshape ADMX} Collaboration, S.~J. Asztalos et~al., \emph{{Experimental
  constraints on the axion dark matter halo density}},
  \href{https://doi.org/10.1086/341130}{\emph{Astrophys. J. Lett.} {\bfseries
  571} (2002) L27} [\href{https://arxiv.org/abs/astro-ph/0104200}{{\ttfamily
  astro-ph/0104200}}].

\bibitem{PhysRevD.69.011101}
S.~J. Asztalos, R.~F. Bradley, L.~Duffy, C.~Hagmann, D.~Kinion, D.~M. Moltz,
  L.~J. Rosenberg, P.~Sikivie, W.~Stoeffl, N.~S. Sullivan, D.~B. Tanner, K.~van
  Bibber and D.~B. Yu, \emph{Improved rf cavity search for halo axions},
  \href{https://doi.org/10.1103/PhysRevD.69.011101}{\emph{Phys. Rev. D}
  {\bfseries 69} (2004) 011101}.

\bibitem{Du:2018uak}
{\scshape ADMX} Collaboration, N.~Du et~al., \emph{{A Search for Invisible
  Axion Dark Matter with the Axion Dark Matter Experiment}},
  \href{https://doi.org/10.1103/PhysRevLett.120.151301}{\emph{Phys. Rev. Lett.}
  {\bfseries 120} (2018) 151301}
  [\href{https://arxiv.org/abs/1804.05750}{{\ttfamily 1804.05750}}].

\bibitem{Piffl_2014}
T.~Piffl, C.~Scannapieco, J.~Binney, M.~Steinmetz, R.-D. Scholz, M.~E.~K.
  Williams, R.~S. de~Jong, G.~Kordopatis, G.~Matijevič, O.~Bienaymé and
  et~al., \emph{The rave survey: the galactic escape speed and the mass of the
  milky way},
  \href{https://doi.org/10.1051/0004-6361/201322531}{\emph{Astronomy \&
  Astrophysics} {\bfseries 562} (2014) A91}.

\bibitem{Monari_2018}
G.~Monari, B.~Famaey, I.~Carrillo, T.~Piffl, M.~Steinmetz, R.~F.~G. Wyse,
  F.~Anders, C.~Chiappini and K.~Janßen, \emph{The escape speed curve of the
  galaxy obtained from gaia dr2 implies a heavy milky way},
  \href{https://doi.org/10.1051/0004-6361/201833748}{\emph{Astronomy \&
  Astrophysics} {\bfseries 616} (2018) L9}.

\bibitem{Deason_2019}
A.~J. Deason, A.~Fattahi, V.~Belokurov, N.~W. Evans, R.~J.~J. Grand,
  F.~Marinacci and R.~Pakmor, \emph{The local high-velocity tail and the
  galactic escape speed},
  \href{https://doi.org/10.1093/mnras/stz623}{\emph{Monthly Notices of the
  Royal Astronomical Society} {\bfseries 485} (2019) 3514–3526}.

\bibitem{Ling_2004}
F.-S. Ling, P.~Sikivie and S.~Wick, \emph{Diurnal and annual modulation of cold
  dark matter signals},
  \href{https://doi.org/10.1103/physrevd.70.123503}{\emph{Physical Review D}
  {\bfseries 70} (2004) }.

\bibitem{Vergados_2017}
J.~Vergados and Y.~Semertzidis, \emph{Axionic dark matter signatures in various
  halo models},
  \href{https://doi.org/10.1016/j.nuclphysb.2016.12.002}{\emph{Nuclear Physics
  B} {\bfseries 915} (2017) 10–18}.

\bibitem{OHare:2017yze}
C.~A.~J. O'Hare and A.~M. Green, \emph{{Axion astronomy with microwave cavity
  experiments}}, \href{https://doi.org/10.1103/PhysRevD.95.063017}{\emph{Phys.
  Rev. D} {\bfseries 95} (2017) 063017}
  [\href{https://arxiv.org/abs/1701.03118}{{\ttfamily 1701.03118}}].

\bibitem{Foster:2017hbq}
J.~W. Foster, N.~L. Rodd and B.~R. Safdi, \emph{{Revealing the Dark Matter Halo
  with Axion Direct Detection}},
  \href{https://doi.org/10.1103/PhysRevD.97.123006}{\emph{Phys. Rev. D}
  {\bfseries 97} (2018) 123006}
  [\href{https://arxiv.org/abs/1711.10489}{{\ttfamily 1711.10489}}].

\bibitem{OHare:2018trr}
C.~A. O'Hare, C.~McCabe, N.~W. Evans, G.~Myeong and V.~Belokurov, \emph{{Dark
  matter hurricane: Measuring the S1 stream with dark matter detectors}},
  \href{https://doi.org/10.1103/PhysRevD.98.103006}{\emph{Phys. Rev. D}
  {\bfseries 98} (2018) 103006}
  [\href{https://arxiv.org/abs/1807.09004}{{\ttfamily 1807.09004}}].

\bibitem{Bozorgnia:2011tk}
N.~Bozorgnia, G.~B. Gelmini and P.~Gondolo, \emph{{Daily modulation due to
  channeling in direct dark matter crystalline detectors}},
  \href{https://doi.org/10.1103/PhysRevD.84.023516}{\emph{Phys. Rev. D}
  {\bfseries 84} (2011) 023516}
  [\href{https://arxiv.org/abs/1101.2876}{{\ttfamily 1101.2876}}].

\bibitem{Mayet:2016zxu}
F.~Mayet et~al., \emph{{A review of the discovery reach of directional Dark
  Matter detection}},
  \href{https://doi.org/10.1016/j.physrep.2016.02.007}{\emph{Phys. Rept.}
  {\bfseries 627} (2016) 1} [\href{https://arxiv.org/abs/1602.03781}{{\ttfamily
  1602.03781}}].

\bibitem{Caputo:2020quz}
A.~Caputo, A.~J. Millar and E.~Vitagliano, \emph{{Revisiting longitudinal
  plasmon-axion conversion in external magnetic fields}},
  \href{https://doi.org/10.1103/PhysRevD.101.123004}{\emph{Phys. Rev. D}
  {\bfseries 101} (2020) 123004}
  [\href{https://arxiv.org/abs/2005.00078}{{\ttfamily 2005.00078}}].

\bibitem{Weldon:1983jn}
H.~A. Weldon, \emph{{Simple rules for discontinuities in finite temperature
  field theory}}, \href{https://doi.org/10.1103/PhysRevD.28.2007}{\emph{Phys.
  Rev. D} {\bfseries 28} (1983) 2007}.

\bibitem{Kapusta:2006pm}
J.~Kapusta and C.~Gale, \emph{{Finite-temperature field theory: principles and
  applications}}, Cambridge Monographs on Mathematical Physics. Cambridge
  University Press, 2011,
  \href{https://doi.org/10.1017/CBO9780511535130}{10.1017/CBO9780511535130}.

\bibitem{Kuo:1989qe}
T.-K. Kuo and J.~T. Pantaleone, \emph{{Neutrino Oscillations in Matter}},
  \href{https://doi.org/10.1103/RevModPhys.61.937}{\emph{Rev. Mod. Phys.}
  {\bfseries 61} (1989) 937}.

\bibitem{Giunti:2007ry}
C.~Giunti and W.~Chung, \emph{Fundamentals of Neutrino Physics and
  Astrophysics}. 01, 2007,
  \href{https://doi.org/10.1093/acprof:oso/9780198508717.001.0001}{10.1093/acprof:oso/9780198508717.001.0001}.

\bibitem{Hardy:2016kme}
E.~Hardy and R.~Lasenby, \emph{{Stellar cooling bounds on new light particles:
  plasma mixing effects}},
  \href{https://doi.org/10.1007/JHEP02(2017)033}{\emph{JHEP} {\bfseries 02}
  (2017) 033} [\href{https://arxiv.org/abs/1611.05852}{{\ttfamily
  1611.05852}}].

\end{thebibliography}\endgroup
 
\end{document}